\documentclass[reviewcopy]{elsarticle}

\usepackage[reviewcopy]{adndt}
\usepackage{longtable}
\usepackage{pdflscape} 
\usepackage{subfig}
\usepackage{amsmath}
\usepackage{rotating}
\usepackage{lineno}

\usepackage{amsmath}

\usepackage{amssymb}

\biboptions{square,sort&compress}
\bibpunct[]{[}{]}{,}{n}{}{;}
\begin{document}

\begin{frontmatter}

\title{Comprehensive Review of 2$\beta$ Decay Half-Lives}
\cortext[cor1]{Corresponding author}
\author[label1]{B. Pritychenko\corref{cor1}}
\ead{pritychenko@bnl.gov}
\address[label1]{National Nuclear Data Center, Brookhaven National Laboratory, \\ Upton, NY 11973-5000, USA}

\author[label2,label3]{V.I. Tretyak}
\address[label2]{Institute for Nuclear Research, Kyiv, 03028, Ukraine}
\address[label3]{INFN, Laboratori Nazionali del Gran Sasso, 67100 Assergi (AQ), Italy}

\date{\today}

\begin{abstract}
The double-beta (2$\beta$)-decay is the rarest nuclear physics process, and its experimental half-lives (T$_{1/2}$) exceed the age of the Universe from nine to fourteen orders of magnitude. Double-beta decay was observed, and its half-life was measured in 14 parent nuclei using direct, radiochemical, and geochemical methods. 
The decay observables are analyzed using the Evaluated Nuclear Structure Data File (ENSDF) procedures, 
and the recommended T$_{1/2}$ were deduced. Using the calculated values of phase factors,  the effective nuclear matrix elements were extracted and compared with available data. Thousands of theoretical and experimental works have been dedicated to these topics in the last 85 years, and we present two data sets of recommended values to encapsulate the results.
\end{abstract}

\begin{keyword}
$2\beta$-decay, half-lives, data compilation, and evaluation
\end{keyword}

\end{frontmatter}




\newpage

\tableofcontents
\listofDtables
\listoffigures
\vskip5pc

\section{Introduction}

Double-beta decay was originally introduced by Maria Goeppert-Mayer~\cite{Goe35} (following the suggestion of Eugene Wigner) as a nuclear disintegration of a parent nucleus with simultaneous emission of  two electrons and two neutrinos, and change of the atomic number of a daughter nucleus  by two units 
\begin{equation}
\label{myeq.2b}  
(Z,A) \rightarrow (Z+2,A) + 2 e^{-} +  2\bar{\nu}_{e}
\end{equation}

\begin{figure}
\centering
\includegraphics[width=0.8\textwidth]{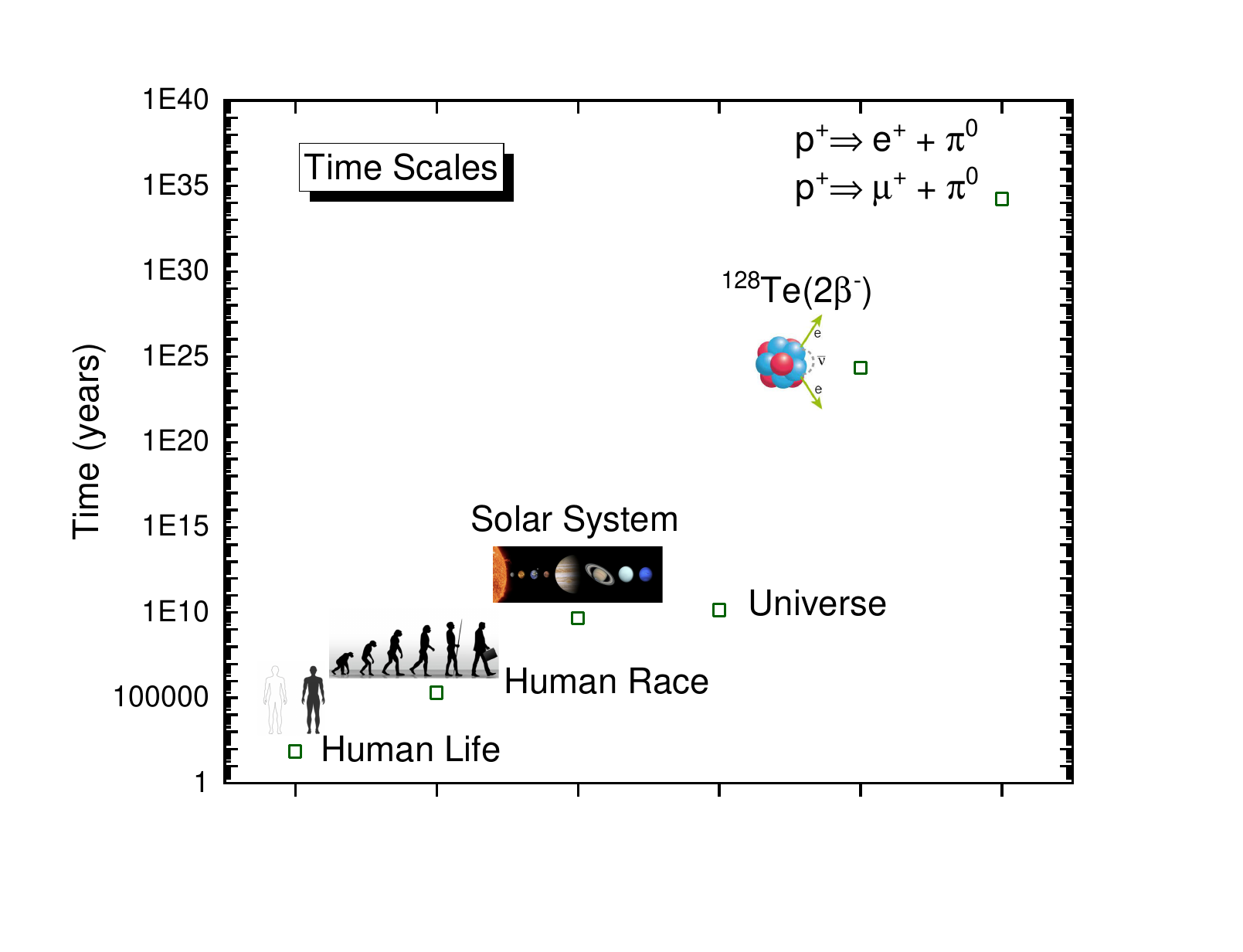}
\caption{Human/Astrophysical/Nuclear Time Scales. The nuclear time scale is based on the $^{128}$Te(2$\beta^{-}$) half-life and the lower limit for a hypothetical proton decay.}
\label{fig:Time}
\end{figure}

 There are four possible double-beta decay processes: $2 \beta^{-}$,  $2 \beta^{+}$, $\epsilon$$\beta^{+}$, 2$\epsilon$ and several major decay modes: 
 two-neutrino (2$\nu$), neutrinoless (0$\nu$) and Majoron emission (0$\nu J$).  
  2$\nu$-mode is not prohibited by conservation laws and occurs as a second-order process compared to the regular $\beta$-decay.  Due to the rare nature of this process and the impact of radioactive background, stable  (or, more precisely, long-lived) nuclei are frequently used in double-beta decay investigations.   The 2$\nu$ and (2$\nu$ + 0$\nu$) decays were observed in 14 parent nuclei employing geochemical, radiochemical, and direct measurements~\cite{Tre95,Tre02}, and the deduced half-lives exceed the age of the Universe~\cite{Goh21} by many orders of magnitude. Figure~\ref{fig:Time} data illustrate that double-beta decay half-lives exceed human and astrophysical periods, and only the hypothetical decay of protons into other subatomic particles~\cite{Sak67} is rarer. 
   
Presently, 2$\beta (2\nu)$-decay is the rarest nuclear decay in nature.  0$\nu$-mode differs from the 2$\nu$-mode because no neutrinos are emitted during the decay. 
This implies that the lepton number is not conserved and a neutrino is a massive Majorana particle (a particle that equals its anti-particle). The quest for neutrinoless decay has generated a lot of excitement in the research community since the early beginnings~\cite{Tre11}, experiments evolved over the years, and limits were improved, however, the neutrinoless mode remains elusive. The current data show that the elusive mode half-life limits are four to five orders of magnitude higher than two-neutrino mode half-lives. It may take many years before the neutrinoless decay will be found or proven as nonexistent while two-neutrino decay data are available and require archiving efforts.

As of today, there are multiple evaluation efforts~\cite{Ell02,Pri14,PDG,Bar02,Bar06,Bar09,Bar10,ASBar10,Bar13,Bar15,Bar19,Bar20} based on different techniques and one comprehensive compilation~\cite{Tre95,Tre02} of double-beta decay data. Multiple evaluation efforts may produce variant, time-dependent results and confuse users. To resolve this issue, we combine comprehensively compiled data from the Institute for Nuclear Research of NASU, Kyiv, Ukraine~\cite{Tre95,Tre02,Tre11} and NNDC~\cite{Pri24} with the best ENSDF evaluation practices and procedures developed by the international Nuclear Structure and Decay Data (NSDD) evaluators network~\cite{ensdf,Dim20,NSDD}. The decay data evaluation methods, techniques, and recommended values will be given in the subsequent sections.

\section{2$\beta$-Decay Measurements and Observables}

The 2$\beta(2\nu$)-decay has been observed in many nuclei from $Z$=20 to $Z$=92, and its experimental T$_{1/2}^{2\nu}$ values have been reported. The decay observables include electrons, positrons, 511 keV $\gamma$-rays (e.g.~\cite{Sae94}), X-rays, and chemically separated daughter atoms. 
These observations can be separated into two categories: direct (decay products) and geochemical/radiochemical (daughter nuclei)
\begin{itemize}
\item Direct Measurements
\begin{itemize}
\item Observation of two-electron events (2$e^{-}$)
\item Fitting residual $\beta$-spectra with calculated two-neutrino mode distributions (2$\beta$Fit)
\item Background subtraction with natural and enriched samples (BackSub)
\item Registration of X-rays and conversion electrons in daughter nuclei (Daughter X-rays)
\item Observations of $\gamma$-transitions in daughter nuclei (Daughter $\gamma$-rays)
\end{itemize}
\item Geochemical and Radiochemical Measurements
\begin{itemize}
\item Detection and counting of daughter nuclei in samples 
\end{itemize}
\end{itemize}

The precise measurements of 2$\beta$-decay are difficult because of the large mass (number of parent nuclei), good energy~\cite{Tre95}, and time resolutions~\cite{2012CH52} are needed.  2$\beta$-decay is a rare process that requires large numbers of studied nuclei in a detector active volume. Several of the current detectors are compliant with the requirements mentioned earlier. Time projection chambers (TPC) provide extensive information on two-electron energy and angular distributions while HPGe (High Purity Germanium) and bolometric scintillator calorimeters combine large volumes of parent nuclei with good energy resolutions. Large detector masses increase overall sensitivity towards neutrinoless events and provide residual $\beta$-spectra for two-neutrino events. 

The $\beta$-spectra analysis could be less reliable for the deduction of absolute values due to insufficient rejection of background contributions. The current situation could be summarized using logic and mathematics regarding necessity and sufficiency. Thus, the constantly evolving $^{76}$Ge two-neutrino half-lives show that the observation of residual background is necessary but insufficient for deducing precise T$_{1/2}$ values. In brief, detectors that accommodate all three requirements are needed for precise measurements.  Unfortunately, such detectors are not available, and it is a data evaluator task to interpret results from many imperfect or discrepant measurements correctly.

\section{Data Compilations}

Nuclear data compilations create a foundation for data evaluations. 

\subsection{Compiled Data}
The 2$\beta$-decay data, in the present work, were compiled capitalizing on the best practices of the Kyiv and the USNDP groups~\cite{Tre02,ensdf} using the data program nuclear bibliography databases~\cite{Pri11,Zer22}. All relevant publications before January 1, 2024 (literature time cut) were analyzed for final results on half-lives and experimental techniques and summarized in Table~\ref{table1}.
\begin{longtable}[c]{l c| c| c| c } 
\caption{Compilation of experimental half-lives for $2 \beta$(2$\nu$)-decay ordered by year ascending. Ground-to-ground state transitions are assumed unless daughter nucleus  $\gamma$-rays were measured. Ge-data and their re-analyses are grouped by experiment. Statistical and systematic uncertainties from the original works are added in quadrature.} \\
\hline\hline
   \textbf{Nuclide}  &   \textbf{Decay} &  \textbf{T$_{1/2}^{2 \nu}$(y)} &  \textbf{Method}  &  \textbf{Reference}  \\
 \hline 
\endfirsthead
\hline
   \textbf{Nuclide}  &   \textbf{Decay} &  \textbf{T$_{1/2}^{2 \nu}$(y)} &  \textbf{Method}  &  \textbf{Reference}   \\
\hline 
\endhead
\hline
\multicolumn{5}{r}{\textit{Continued on next page ...}} \\
\endfoot
\endlastfoot   
$^{48}$Ca & 2$\beta^-$(0$^{+}_{1}$$\rightarrow$0$^{+}_{1}$) & 4300$^{+2788}_{-1780}\times10^{16}$  & 2$e^{-}$ & \cite{1996BA80}   \\
                  &                                                              & 4200$^{+3300}_{-1300}\times10^{16}$  & 2$\beta$Fit (TGV) & \cite{2000BR63}  \\
                  &                                                              & 6400$^{+1389}_{-1082}\times10^{16}$  & 2$e^{-}$ (NEMO-3) & \cite{2016AR19}  \\ 
\hline
$^{76}$Ge & 2$\beta^-$(0$^{+}_{1}$$\rightarrow$0$^{+}_{1}$) & 900$^{+100}_{-100}\times10^{18}$  & BackSub, 2$\beta$Fit &  \cite{1990VA18} \\
\cline{3-5}   
                  &                                                              & 1270$^{+210}_{-160}\times10^{18}$  & 2$\beta$Fit & \cite{1990MI23,1994AV07}  \\
\cline{3-5}                 
                  &                                                              & 840$^{+100}_{-80}\times10^{18}$  & 2$\beta$Fit& \cite{1991AV01,1993BR22}  \\                  
                  &                                                              & 1450$^{+150}_{-150}\times10^{18}$ & 2$\beta$Fit (IGEX) & \cite{1996AA02,1999MO40}  \\
\cline{3-5}                   
                  &                                                              & 1550$^{+190}_{-150}\times10^{18}$  & 2$\beta$Fit (Heidelberg-Moscow) & \cite{1997GU13,2001KL11}  \\ 
                  &                                                              & 1740$^{+180}_{-160}\times10^{18}$  & 2$\beta$Fit (Heidelberg-Moscow) & \cite{2003DO12}  \\                                     
                  &                                                              & 1780$^{+71}_{-91}\times10^{18}$  & 2$\beta$Fit (Heidelberg-Moscow) & \cite{2005BA60}  \\ 
\cline{3-5}                  
                  &                                                              & 1926$^{+94}_{-94}\times10^{18}$  & 2$\beta$Fit (GERDA-I) & \cite{2013AG02,2015AG06}  \\ 
\cline{3-5}                  
                  &                                                              & 2022$^{+42}_{-42}\times10^{18}$  & 2$\beta$Fit (GERDA) & \cite{2020AGP,2023AG05}  \\ 
\hline
$^{82}$Se & 2$\beta^-$(0$^{+}_{1}$$\rightarrow$0$^{+}_{1}$) & 600$^{+600}_{-300}\times10^{17}$  & GeoChemistry &  \cite{1967KI04} \\ 
                  &                                                              & 1400$^{+280}_{-280}\times10^{17}$  & GeoChemistry &  \cite{1969KI17} \\     
                  &                                                              & 2760$^{+880}_{-880}\times10^{17}$  & GeoChemistry &  \cite{1973SR05} \\ 
                  &                                                              & 1450$^{+150}_{-150}\times10^{17}$  & GeoChemistry &  \cite{1983KIZV} \\   
                  &                                                              & 1030$^{+330}_{-420}\times10^{17}$  & GeoChemistry &  \cite{1985MA57,1987MU18} \\  
                  &                                                              & 1000$^{+400}_{-400}\times10^{17}$  & GeoChemistry &  \cite{1986MAYT} \\  
                  &                                                              & 1300$^{+50}_{-50}\times10^{17}$  & GeoChemistry &  \cite{1986KIZQ} \\    
                  &                                                              & $<$1400$^{+200}_{-200}\times10^{17}$  & GeoChemistry &  \cite{1986LI10} \\       
                  &                                                              & 1200$^{+100}_{-100}\times10^{17}$  & GeoChemistry &  \cite{1988LI11} \\    
                  &                                                              & 1000$^{+0}_{-0}\times10^{17}$  & GeoChemistry &  \cite{1991MA64} \\  
                  &                                                              & 1080$^{+260}_{-60}\times10^{17}$  & 2$e^{-}$ &  \cite{1987EL11,1992EL07} \\   
                  &                                                              & 830$^{+122}_{-122}\times10^{17}$  & 2$e^{-}$ (NEMO-2) &  \cite{1998AR10} \\ 
                  &                                                              & 939$^{+60}_{-60}\times10^{17}$  & 2$e^{-}$ (NEMO-3) &  \cite{2005AR27,2018AR06} \\   
                  &                                                              & 860$^{+19}_{-13}\times10^{17}$  &  2$\beta$Fit (CUPID-0) &  \cite{2019AZ04} \\  
                  &                                                              & 869$^{+10}_{-8}\times10^{17}$  &  2$\beta$Fit (CUPID-0+) &  \cite{2023AZ05} \\ 
\hline
$^{78}$Kr & 2$K$(0$^{+}_{1}$$\rightarrow$0$^{+}_{1}$) & 190$^{+133}_{-76}\times10^{20}$  & Daughter X-rays &  \cite{2013GA13,2017RA27} \\  
\hline
$^{96}$Zr & 2$\beta^-$(0$^{+}_{1}$$\rightarrow$0$^{+}_{1}$) & 390$^{+90}_{-90}\times10^{17}$  & GeoChemistry &  \cite{1993KA12} \\ 
                  &                                                              & 210$^{+82}_{-45}\times10^{17}$  & 2$e^{-}$ (NEMO-2) &  \cite{1999AR25} \\
                  &                                                              & 94$^{+32}_{-32}\times10^{17}$  & GeoChemistry &  \cite{2001WI17} \\  
                  &                                                              & 235$^{+21}_{-21}\times10^{17}$  & 2$e^{-}$ (NEMO-3) &  \cite{2010AR07} \\ 
                  &                                                              & 203$^{+46}_{-31}\times10^{17}$  & GeoChemistry &  \cite{2018MA51} \\  
\hline                                                                                                                                                                                                                                                                                                      
$^{100}$Mo & 2$\beta^-$(0$^{+}_{1}$$\rightarrow$0$^{+}_{1}$) & 330$^{+200}_{-100}\times10^{16}$  & 2$e^{-}$ &  \cite{1990VA10} \\ 
		   &                                                            & 1150$^{+300}_{-200}\times10^{16}$  & 2$\beta$Fit &  \cite{1991EJ02} \\ 
		   &                                                            & 950$^{+98}_{-98}\times10^{16}$  & 2$e^{-}$ (NEMO-2) &  \cite{1995DA37} \\ 	
		   &                                                            & 760$^{+220}_{-140}\times10^{16}$  & 2$\beta$Fit &  \cite{1997AL02} \\ 	
		   &                                                            & 682$^{+78}_{-86}\times10^{16}$  & 2$e^{-}$ &  \cite{1991EL04,1997DE40} \\ 	
		   &                                                            & 720$^{+201}_{-201}\times10^{16}$  & 2$\beta$Fit &  \cite{2001AS06} \\ 
		   &                                                            & 210$^{+30}_{-30}\times10^{16}$  & GeoChemistry &  \cite{2004HI19} \\ 
		   &                                                            & 715$^{+76}_{-76}\times10^{16}$  & 2$\beta$Fit &  \cite{2014CA46} \\ 		   
		   &                                                            & 690$^{+40}_{-40}\times10^{16}$  & 2$\beta$Fit &  \cite{2017AR18} \\ 
		   &                                                            & 681$^{+38}_{-40}\times10^{16}$  & 2$e^{-}$ (NEMO-3) &  \cite{2005AR27,2019AR04} \\ 	
		   &                                                            & 707$^{+11}_{-11}\times10^{16}$  & 2$e^{-}$ (CUPID-Mo) &  \cite{2020AR09,2023AU05} \\ 			   
\hline  
$^{100}$Mo & 2$\beta^-$(0$^{+}_{1}$$\rightarrow$0$^{+}_{2}$) & 610$^{+180}_{-110}\times10^{18}$  & Daughter $\gamma$-rays &  \cite{1995BA29} \\ 
                    &                                                            & 930$^{+313}_{-220}\times10^{18}$  & Daughter $\gamma$-rays &  \cite{1999BB19} \\ 		   	   		   	   
                    &                                                            & 590$^{+180}_{-125}\times10^{18}$  & Daughter $\gamma$-rays &  \cite{2001DE17} \\ 
                    &                                                            & 570$^{+150}_{-120}\times10^{18}$  & 2$e^{-}$+Daughter $\gamma$-rays (NEMO-3) &  \cite{2007AR02} \\ 
                    &                                                            & 550$^{+120}_{-90}\times10^{18}$  & Daughter $\gamma$-rays (TUNL-ITEP)&  \cite{2006HO17,2009KI04} \\  
                    &                                                            & 690$^{+120}_{-110}\times10^{18}$  & Daughter $\gamma$-rays (ARMONIA) &  \cite{2010BE34} \\    
                    &                                                            & 750$^{+85}_{-85}\times10^{18}$  & Daughter $\gamma$-rays (NEMO-3) &  \cite{2014AR08} \\   
		   &                                                            & 750$^{+89}_{-85}\times10^{18}$  & 2$e^{-}$ (CUPID-Mo) &  \cite{2023AU01} \\  
                    
\hline 
$^{116}$Cd & 2$\beta^-$(0$^{+}_{1}$$\rightarrow$0$^{+}_{1}$) & 260$^{+90}_{-50}\times10^{17}$  & 2$e^{-}$ (ELEGANT V)  &  \cite{1995EJ01} \\   
                    &                                                            & 270$^{+102}_{-72}\times10^{17}$  & 2$\beta$Fit  &  \cite{1995DA09} \\ 	
                    &                                                            & 375$^{+41}_{-41}\times10^{17}$  & 2$e^{-}$ (NEMO-2) &  \cite{1995AR26,1996AR36} \\  
                    &                                                            & 290$^{+36}_{-36}\times10^{17}$  & 2$e^{-}$ (NEMO-2) &  \cite{ASBar10} \\   
                    &                                                            & 290$^{+40}_{-30}\times10^{17}$  & 2$\beta$Fit  &  \cite{2000DA27,2003DA09,2003DA24} \\     
                    &                                                            & 274$^{+18}_{-18}\times10^{17}$  & 2$e^{-}$ (NEMO-3) &  \cite{2017AR01} \\     
                    &                                                            & 263$^{+11}_{-12}\times10^{17}$  & 2$\beta$Fit (Aurora) &  \cite{2018BA44} \\               
\hline 
$^{128}$Te & 2$\beta^-$(0$^{+}_{1}$$\rightarrow$0$^{+}_{1}$) & 154$^{+17}_{-17}\times10^{22}$  & GeoChemistry  &  \cite{1975HE04} \\      
                   &                                                            & $>$800$^{+0}_{-0}\times10^{22}$  & GeoChemistry  &  \cite{1983KI02,1983KI03} \\ 
                   &                                                            & 140$^{+40}_{-40}\times10^{22}$  & GeoChemistry  &  \cite{1986MAYT} \\    
                   &                                                            & $>$500$^{+0}_{-0}\times10^{22}$  & GeoChemistry  &  \cite{1986KIZQ} \\ 
                   &                                                            & 180$^{+70}_{-70}\times10^{22}$  & GeoChemistry  &  \cite{1988LI12} \\ 
                   &                                                            & 200$^{+0}_{-0}\times10^{22}$  & GeoChemistry  &  \cite{1991MA64} \\   
                   &                                                            & 770$^{+40}_{-40}\times10^{22}$  & GeoChemistry  &  \cite{1992BE30,1993BE04} \\   
                   &                                                            & 720$^{+30}_{-30}\times10^{22}$  & GeoChemistry  &  \cite{1993Da19} \\     
                   &                                                            & 220$^{+30}_{-30}\times10^{22}$  & GeoChemistry  &  \cite{1996TA04} \\       
                   &                                                            & 241$^{+39}_{-39}\times10^{22}$  & GeoChemistry  &  \cite{2008ME10} \\  
                   &                                                            & 230$^{+30}_{-30}\times10^{22}$  & GeoChemistry  &  \cite{2008TH07} \\ 
                   &                                                            & 220$^{+20}_{-20}\times10^{22}$  & GeoChemistry  &  \cite{2017MEZS} \\       
\hline 
$^{130}$Te & 2$\beta^-$(0$^{+}_{1}$$\rightarrow$0$^{+}_{1}$) & 1400$^{+0}_{-0}\times10^{18}$  & GeoChemistry  &  \cite{1950IN03} \\ 
                   &                                                           & 820$^{+64}_{-64}\times10^{18}$  & GeoChemistry  &  \cite{1966TA02} \\  
                   &                                                           & 600$^{+600}_{-300}\times10^{18}$  & GeoChemistry  &  \cite{1967KI04} \\    
                   &                                                           & 300$^{+40}_{-40}\times10^{18}$  & GeoChemistry  &  \cite{1967GE12} \\    
                   &                                                           & 2188$^{+696}_{-528}\times10^{18}$  & GeoChemistry  &  \cite{1968KI02} \\   
                   &                                                           & 2030$^{+300}_{-300}\times10^{18}$  & GeoChemistry  &  \cite{1969AL22} \\  
                   &                                                           & 2510$^{+146}_{-146}\times10^{18}$  & GeoChemistry  &  \cite{1972SR03} \\    
                   &                                                           & 2830$^{+300}_{-300}\times10^{18}$  & GeoChemistry  &  \cite{1972SR06} \\   
                   &                                                           & 1031$^{+51}_{-51}\times10^{18}$  & GeoChemistry  &  \cite{1975HE04} \\    
                   &                                                           & 2600$^{+280}_{-280}\times10^{18}$  & GeoChemistry  &  \cite{1983KI02,1983KI03} \\    
                   &                                                           & 2550$^{+200}_{-200}\times10^{18}$  & GeoChemistry  &  \cite{1983KIZV} \\    
                   &                                                           & 700$^{+200}_{-200}\times10^{18}$  & GeoChemistry  &  \cite{1986MAYT} \\  
                   &                                                           & 2125$^{+625}_{-625}\times10^{18}$  & GeoChemistry  &  \cite{1986KIZQ} \\      
                   &                                                           & 1000$^{+300}_{-300}\times10^{18}$  & GeoChemistry  &  \cite{1986RI02} \\    
                   &                                                           & $<$1250$^{+80}_{-80}\times10^{18}$  & GeoChemistry  &  \cite{1986LI10} \\        
                   &                                                           & 750$^{+300}_{-300}\times10^{18}$  & GeoChemistry  &  \cite{1988LI11} \\       
                   &                                                           & 800$^{+0}_{-0}\times10^{18}$  & GeoChemistry  &  \cite{1991MA64} \\   
                   &                                                           & 2700$^{+100}_{-100}\times10^{18}$  & GeoChemistry  &  \cite{1992BE30,1993BE04} \\   
                   &                                                           & 790$^{+100}_{-100}\times10^{18}$  & GeoChemistry  &  \cite{1996TA04} \\        
                   &                                                           & 610$^{+322}_{-377}\times10^{18}$  & 2$\beta$Fit   &  \cite{2003AR02} \\  
                   &                                                           & 900$^{+140}_{-140}\times10^{18}$  & GeoChemistry  &  \cite{2008ME10} \\   
                   &                                                           & 800$^{+110}_{-110}\times10^{18}$  & GeoChemistry  &  \cite{2008TH07} \\    
                   &                                                           & 700$^{+142}_{-142}\times10^{18}$  & 2$e^{-}$ (NEMO-3)  &  \cite{2011AR09} \\  
                   &                                                           & 820$^{+63}_{-63}\times10^{18}$  & 2$\beta$Fit (CUORE)  &  \cite{2017AL46} \\     
                   &                                                            & 820$^{+70}_{-70}\times10^{22}$  & GeoChemistry  &  \cite{2017MEZS} \\       
                   &                                                           & 876$^{+17}_{-18}\times10^{18}$  & 2$\beta$Fit (CUORE)  &  \cite{2019CA21,2021AD18,2021AD18E} \\  
\hline  
$^{124}$Xe & 2$K$(0$^{+}_{1}$$\rightarrow$0$^{+}_{1}$) & 180$^{+51}_{-51}\times10^{20}$  & 2$\beta$Fit  (XENON1T) &  \cite{2019AP03} \\
                  &  2$\epsilon$(0$^{+}_{1}$$\rightarrow$0$^{+}_{1}$)  & 118$^{+19}_{-19}\times10^{20} $  & 2$\beta$Fit  (XENONnT) &  \cite{2022AP04} \\ 
                  &  2$\epsilon$(0$^{+}_{1}$$\rightarrow$0$^{+}_{1}$)  & 110$^{+22}_{-22}\times10^{20} $  & 2$\beta$Fit  (XENON1T) &  \cite{2022AP01} \\ 
\hline
$^{136}$Xe & 2$\beta^-$(0$^{+}_{1}$$\rightarrow$0$^{+}_{1}$) & $>$10000$\times10^{18}$  & 2$\beta$Fit   &  \cite{2002BE74} \\  
                   &                                                           & 2110$^{+214}_{-214}\times10^{18}$  & 2$\beta$Fit (EXO)  &  \cite{2011AC03} \\  
                   &                                                           & 2230$^{+221}_{-221}\times10^{18}$  & 2$\beta$Fit (EXO)  &  \cite{2012AU03} \\  
                   &                                                           & 2380$^{+141}_{-141}\times10^{18}$  & 2$\beta$Fit (KamLAND-Zen)   &  \cite{2012GA17} \\ 
                   &                                                           & 2300$^{+122}_{-122}\times10^{18}$  & 2$\beta$Fit (KamLAND-Zen)  &  \cite{2012GA32} \\    
                   &                                                           & 5800$^{+4700}_{-1800}\times10^{18}$  & 2$\beta$Fit   &  \cite{2013GA41} \\   
                   &                                                           & 2165$^{+61}_{-61}\times10^{18}$  & 2$\beta$Fit (EXO)  &  \cite{2014AL03} \\    
                   &                                                           & 2230$^{+76}_{-76}\times10^{18}$  & 2$\beta$Fit  (KamLAND-Zen) &  \cite{2016GA30,2019GA11} \\                     
                   &                                                           & 2340$^{+854}_{-490}\times10^{18}$  & 2$\beta$Fit  (NEXT) &  \cite{2022NO06} \\           
                   &                                                           & 2270$^{+104}_{-104}\times10^{18}$  & 2$\beta$Fit  (PandaX-4T) &  \cite{2022SIZZ} \\           
\hline
$^{130}$Ba & 2$\epsilon$(0$^{+}_{1}$$\rightarrow$0$^{+}_{1}$) & 210$^{+300}_{-80}\times10^{19}$  & GeoChemistry  &  \cite{1996BA24} \\  
                   &                                                          & 220$^{+50}_{-50}\times10^{19}$  & GeoChemistry  &  \cite{2001ME22} \\  
                   &                                                          & 60$^{+11}_{-11}\times10^{19}$  & GeoChemistry  &  \cite{2009PU05} \\   
\hline
$^{150}$Nd & 2$\beta^-$(0$^{+}_{1}$$\rightarrow$0$^{+}_{1}$) & 1880$^{+687}_{-434}\times10^{16}$  & 2$e^{-}$  &  \cite{1995AR08} \\   
                   &                                                          & 675$^{+77}_{-80}\times10^{16}$  & 2$e^{-}$  &  \cite{1997DE40} \\ 
                   &                                                          & 934$^{+66}_{-64}\times10^{16}$  & 2$e^{-}$ (NEMO-3) &  \cite{2009AR10,2016AR12} \\  
\hline
$^{150}$Nd & 2$\beta^-$(0$^{+}_{1}$$\rightarrow$0$^{+}_{2}$) & 133$^{+45}_{-26}\times10^{18}$  & Daughter $\gamma$-rays  &  \cite{2004BA10,2009BA21} \\       
                  &                                                          & 107$^{+46}_{-26}\times10^{18}$  & Daughter $\gamma$-rays  &  \cite{2014KI13} \\   
                 &                                                          & 97$^{+44}_{-34}\times10^{18}$  & Daughter $\gamma$-rays  &  \cite{2021PO16} \\ 
                 &                                                          & 110$^{+26}_{-21}\times10^{18}$  & 2$e^{-}$+Daughter $\gamma$-rays (NEMO-3) &  \cite{2023AG06} \\  
\hline
$^{238}$U & 2$\beta^-$(0$^{+}_{1}$$\rightarrow$0$^{+}_{1}$) & 20$^{+6}_{-6}\times10^{20}$  & RadioChemistry  &  \cite{1991TU02} \\                                                                                                                                                                                                                                                                                                                                                                                                                                                                                                                                                                                                                                                                                                                                                                                                                                                                                                                                                                                     
\hline      
\hline        
\label{table1}
\end{longtable}

\subsection{Compiled Data Analysis}

There is a large number of double-beta decay observations in literature, and it is necessary to apply multiple techniques for data analysis such as the selection of final results for statistical codes and method of visual inspection. For a better understanding of the current situation, we select, plot, and inspect the final results of distinct measurements in chronological order. Experimental half-lives for 14  individual nuclei are shown below. 

\subsubsection{$^{48}$Ca}

\begin{figure}
\centering
\includegraphics[width=0.8\textwidth]{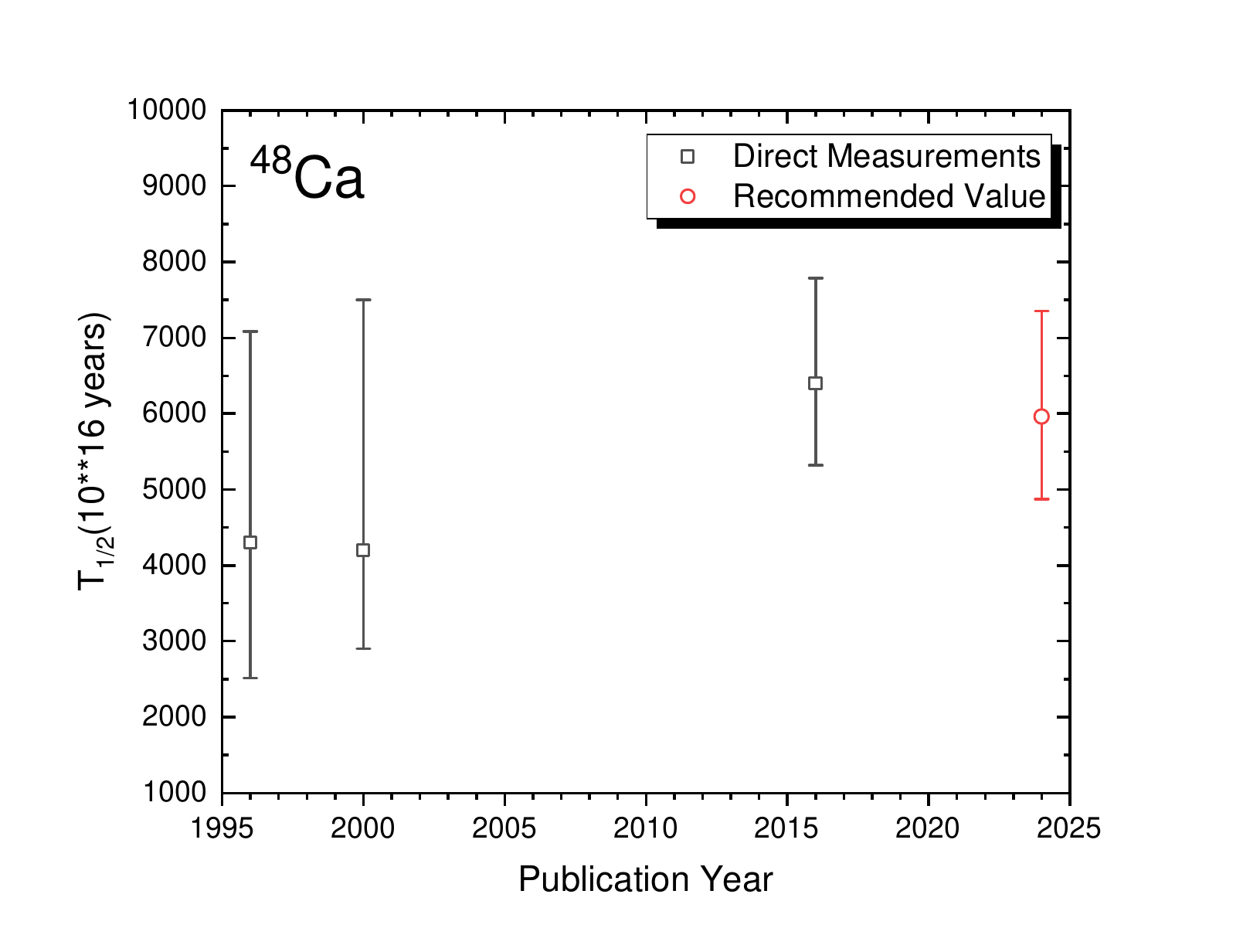}
\caption{$^{48}$Ca experimental half-lives~\cite{1996BA80,2000BR63,2016AR19}.}
\label{fig:48Ca}
\end{figure}

$^{48}$Ca has the highest double-beta transition Q-value among stable even-even nuclei (4.268 MeV~\cite{AME20}), and a very low isotopic abundance of 0.187$\%$~\cite{2016ME20}. The small value of isotopic abundance makes $^{48}$Ca a very expensive material and explains the limited number of measurements.  In this nucleus, a single $\beta$-decay is possible, and advanced detectors like TPC are needed for the separation of one- and two-electron events. The results for 2$\beta^{-}(2\nu)$ transition in $^{48}$Ca~\cite{1996BA80,2000BR63,2016AR19} are graphically depicted in Fig.\ref{fig:48Ca}.

\subsubsection{$^{76}$Ge}

\begin{figure}
\centering
\includegraphics[width=0.8\textwidth]{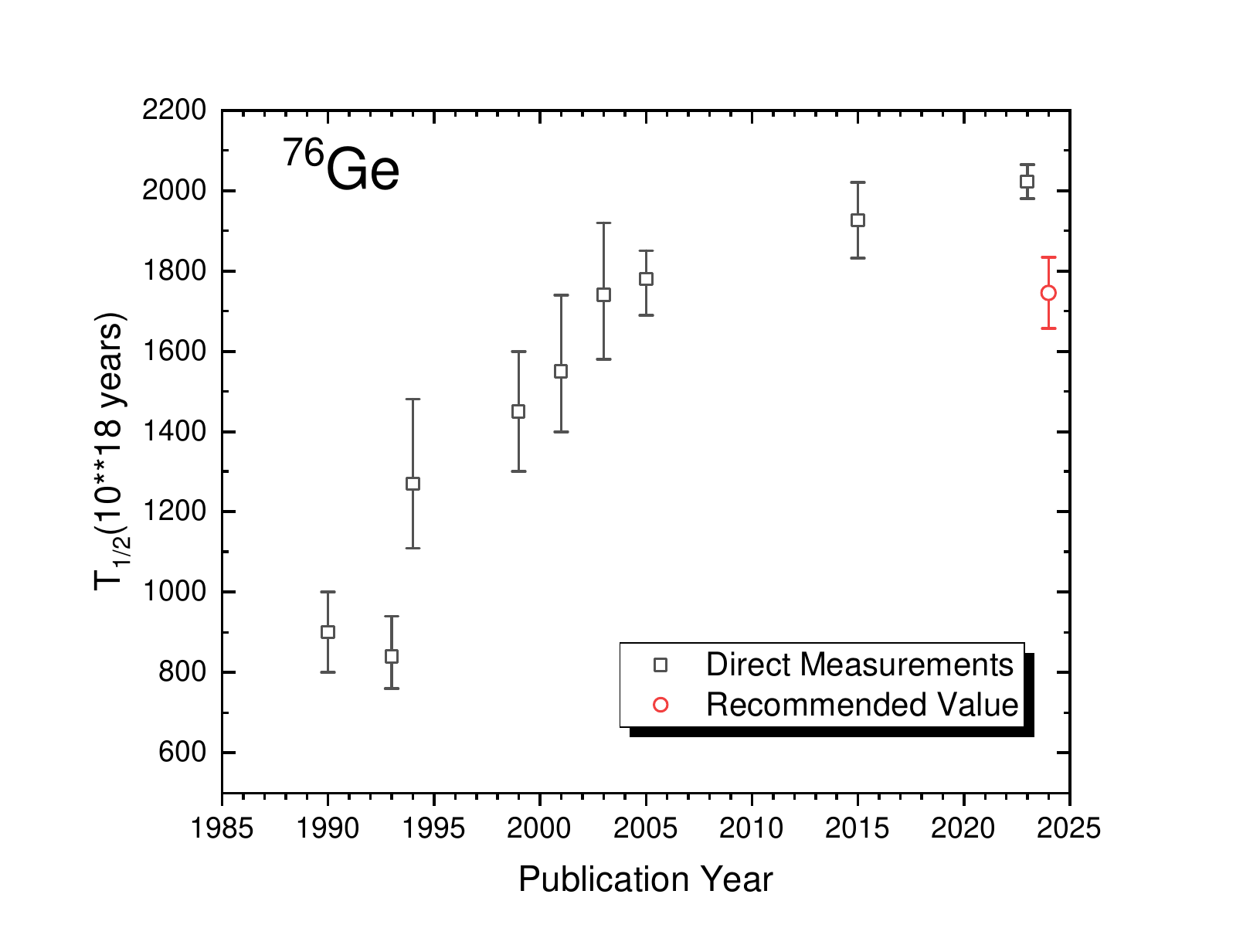}
\caption{$^{76}$Ge experimental half-lives~\cite{1990VA18,1990MI23,1994AV07,1991AV01,1993BR22,1996AA02,1999MO40,1997GU13,2001KL11,2003DO12,2005BA60,2013AG02,2015AG06,2020AGP,2023AG05}.}
\label{fig:76Ge}
\end{figure}

$^{76}$Ge$\rightarrow ^{76}$Se transition is hampered by the intermediate nucleus $^{76}$As ground-state negative parity and relatively low decay Q-value of 2.039 MeV. 
Nevertheless, there are multiple experiments with Ge-detectors because of their excellent energy resolution and detector active volume consisting of  $^{76}$Ge(2$\beta^{-}$) nuclei. Such a combination allows to optimize the effect/background ratio and achieve high sensitivity. The first observation has been made using the combination of enriched to 85$\%$ and natural Ge(Li) crystals located in Yerevan, Armenia salt mine~\cite{1990VA18}, and a surplus counting rate in the enriched detector was attributed to the  2$\beta^-$(0$^{+}_{1}$$\rightarrow$0$^{+}_{1}$)  transitions. The later experiments employed mostly enriched HPGe detectors where the radioactive background was analyzed and subtracted, and the remainder was used to deduce two-neutrino half-lives ($\beta$-spectrum fitting or 2$\beta$Fit).  In all these cases the measured quantity was the total energy deposition,  with no tracking information.  The results for 2$\beta^{-}(2\nu)$ transition in
 $^{76}$Ge~\cite{1990VA18,1990MI23,1994AV07,1991AV01,1993BR22,1996AA02,1999MO40,1997GU13,2001KL11,2003DO12,2005BA60,2013AG02,2015AG06,2020AGP,2023AG05} are graphically depicted in Fig.\ref{fig:76Ge}.

Over the years, the 2$\beta$Fit technique evolved, the radioactive background was reduced and the present-day half-lives are $\sim$2 times higher than the initial claim. A complementary analysis of the Yerevan measurement reveals multiple issues with the background subtraction accuracy and the uncertainty with detector volume measurements~\cite{1990VA18}. In the subsequent years, the enriched Ge detectors were used in other measurements~\cite{1991AV01,1993BR22}. Since Ge-detectors are high-resolution particle calorimeters, only the shapes of the corrected spectra are used to deduce 2$\beta^{-}$ decay. This was achieved with the corrections for $^{210}$Bi background, and no corrections for possible $\gamma$-ray engendered events. Other earlier results~\cite{1990MI23,1994AV07}  were also influenced by the $^{210}$Bi bremsstrahlung and impacted by the variations of radioactive background. These issues explain strong deviations between the Ge results and suggest the selection of post-1995 results for evaluation purposes.

\subsubsection{$^{82}$Se}

\begin{figure}
\centering
\includegraphics[width=0.8\textwidth]{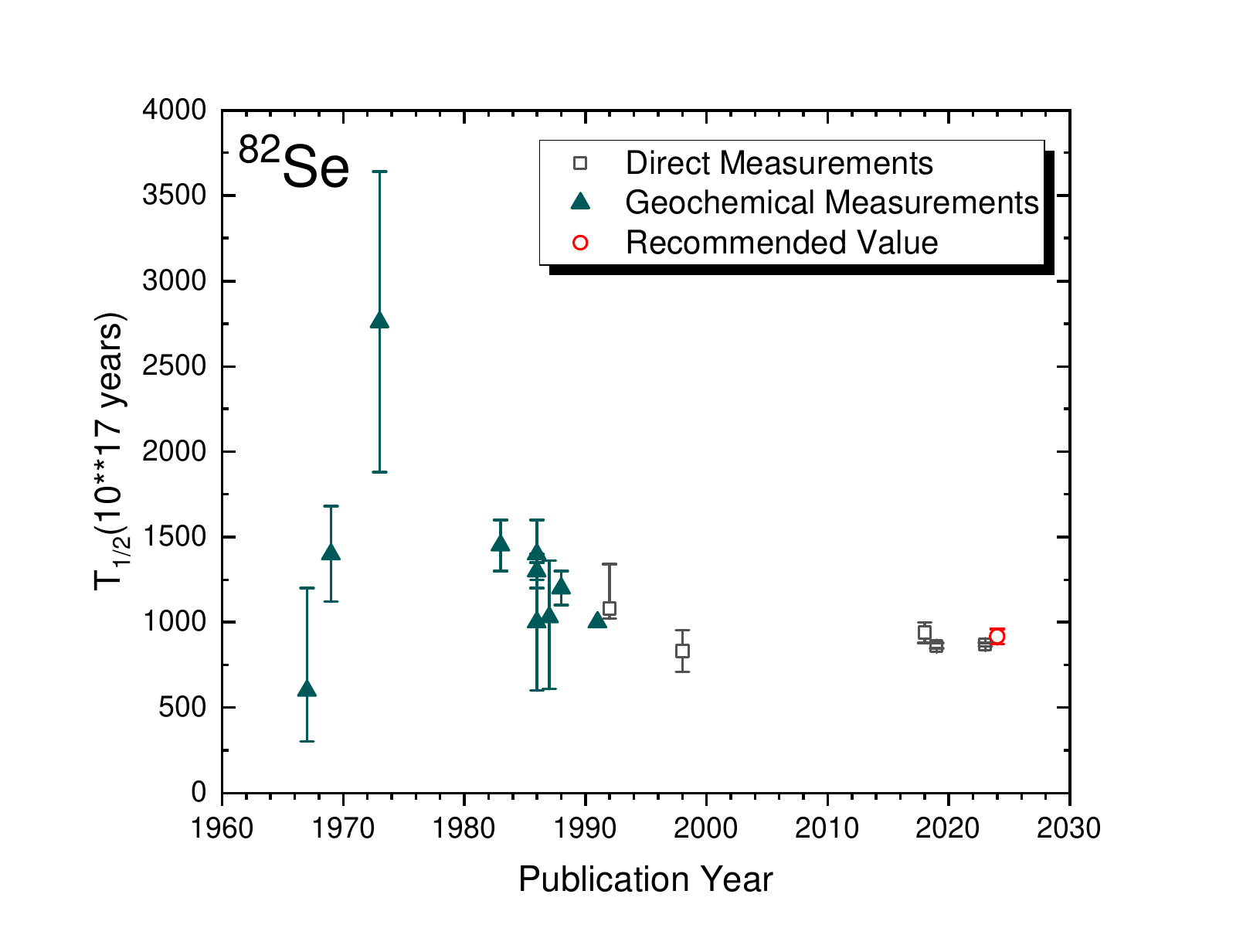}
\caption{$^{82}$Se geochemical~\cite{1967KI04,1969KI17,1973SR05,1983KIZV,1985MA57,1987MU18,1986MAYT,1986KIZQ,1986LI10,1988LI11,1991MA64} and directly measured half-lives~\cite{1987EL11,1992EL07,1998AR10,2005AR27,2018AR06,2019AZ04,2023AZ05}.}
\label{fig:82Se}
\end{figure}

 $^{82}$Se$\rightarrow ^{82}$Kr transition is impeded by the intermediate nucleus $^{82}$Br ground-state negative parity and accelerated by a high decay Q-value of 2.998 MeV that is above the intense lines of natural radioactivity.  Multiple geochemical experiments supplied discrepant results on double-beta decay half-lives in selenium.  This problem was resolved in the 1980s when a two-neutrino mode was directly observed for the first time with UC-Irvine, TPC~\cite{1987EL11,1992EL07}. The observation of two electron events at Irvine provided convincing proof for the existence of double-beta decay. Other measurements were performed in the following years, and the results for 2$\beta^{-}(2\nu)$ transition in $^{82}$Se~\cite{1967KI04,1969KI17,1973SR05,1983KIZV,1985MA57,1987MU18,1986MAYT,1986KIZQ,1986LI10,1988LI11,1991MA64,1987EL11,1992EL07,1998AR10,2005AR27,2018AR06,2019AZ04,2023AZ05} are graphically depicted in Fig.\ref{fig:82Se}. The Figure data show that the recent measurements of NEMO-3 and CUPID-0 collaborations~\cite{2018AR06,2019AZ04,2023AZ05} are compatible with the previous geochemical results in selenium. Finally, the published results of  CUPID-0 collaboration~\cite{2019AZ04} were superseded by the global CUPID-0 background model analysis~\cite{2023AZ05}.

\subsubsection{$^{78}$Kr}
\begin{figure}
\centering
\includegraphics[width=0.8\textwidth]{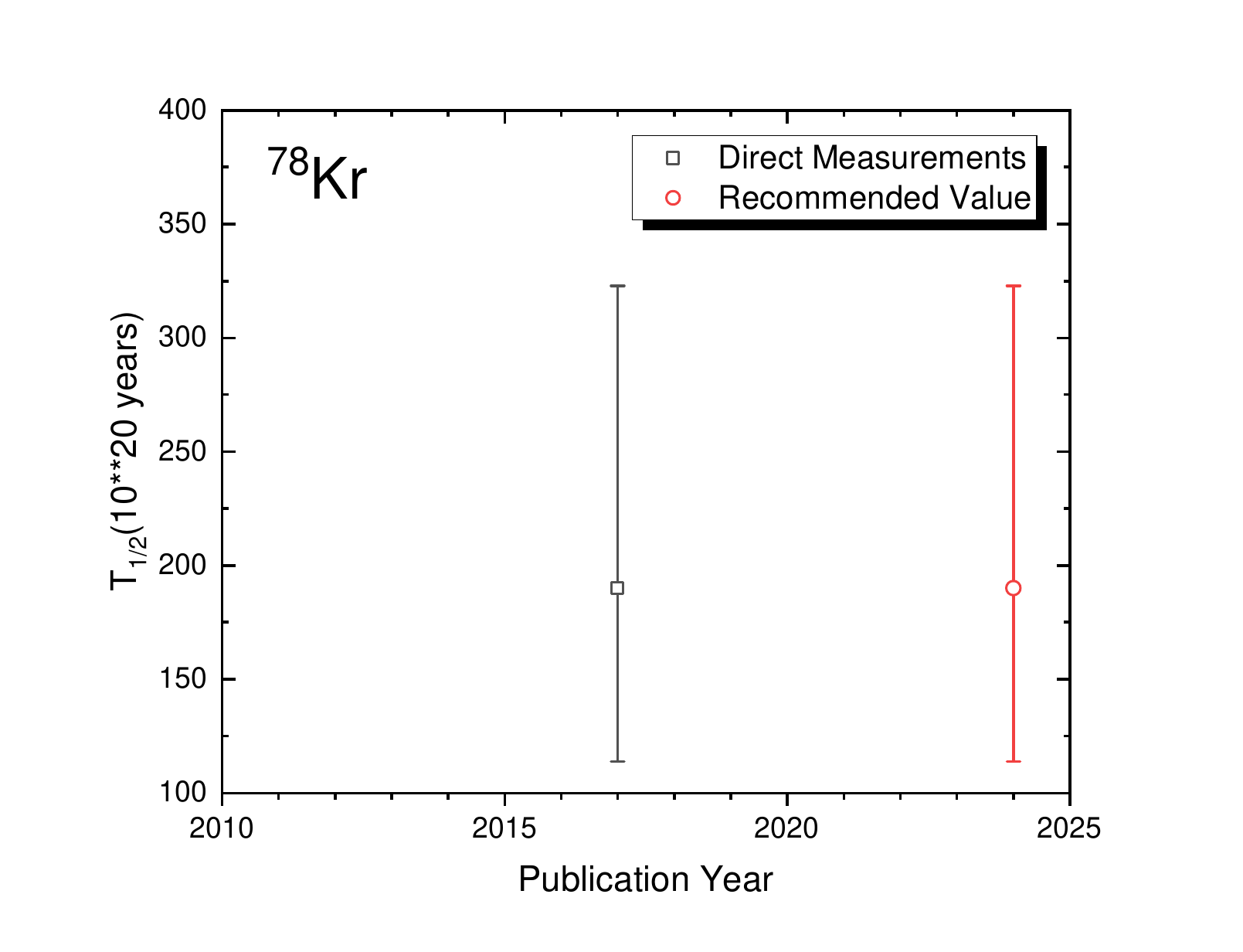}
\caption{$^{78}$Kr experimental half-lives~\cite{2013GA13,2017RA27}.}
\label{fig:78Kr}
\end{figure}

Krypton has the close second largest Q-value for 2$\epsilon$-capture among even-even stable nuclei, and it is used in gaseous detectors. Atomic transitions were used to detect double-electron capture in krypton using the low-background Baksan proportional counters. Enriched samples were needed for such measurements since the isotopic abundance of $^{78}$Kr is 0.36$\%$. The results for 2$\epsilon(2\nu)$ transition in $^{78}$Kr~\cite{2013GA13,2017RA27} are graphically depicted in Fig.\ref{fig:78Kr}.

\subsubsection{$^{96}$Zr}

\begin{figure}
\centering
\includegraphics[width=0.8\textwidth]{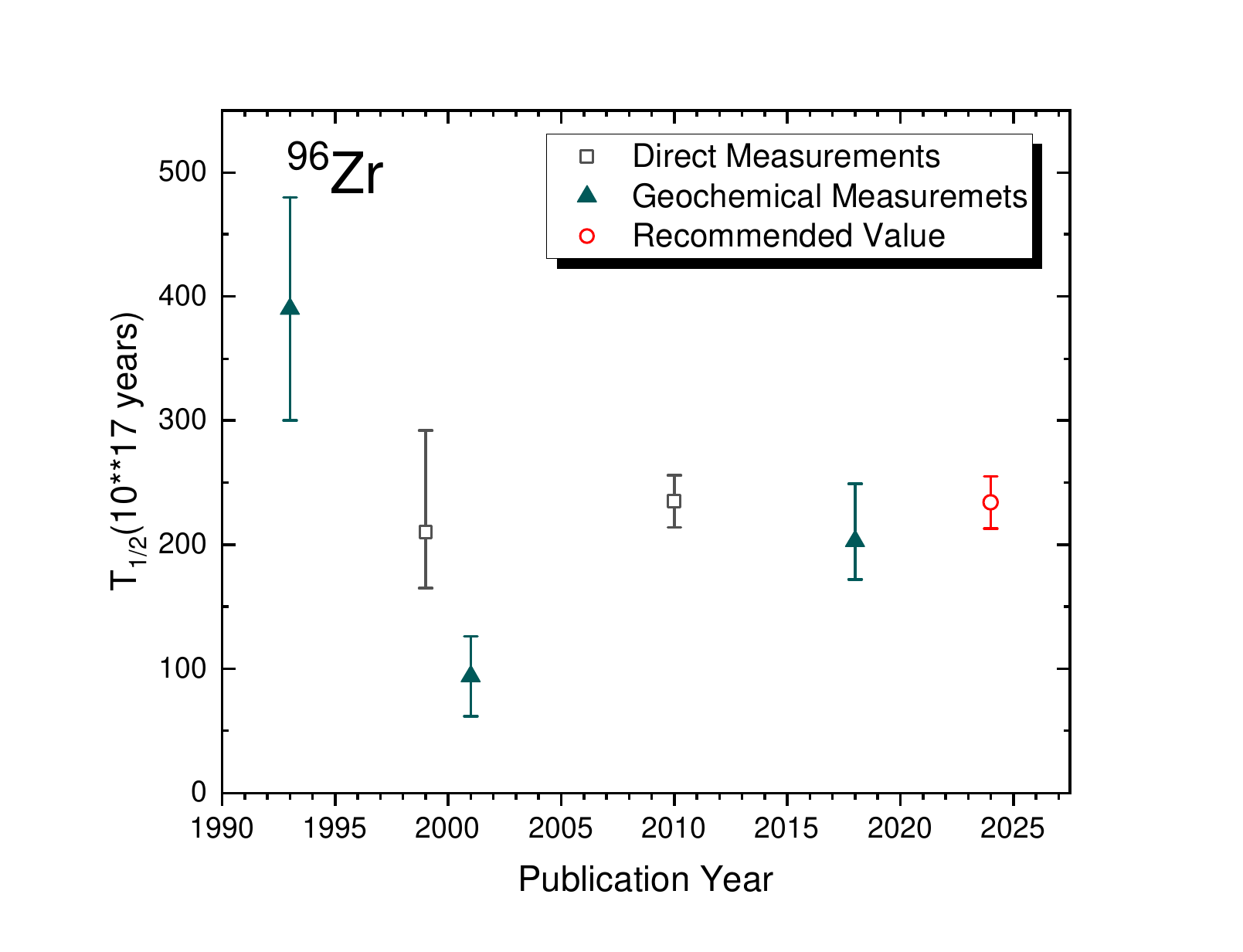}
\caption{$^{96}$Zr geochemical~\cite{1993KA12,2001WI17,2018MA51} and directly measured half-lives~\cite{1999AR25,2010AR07}.}
\label{fig:96Zr}
\end{figure}

Zirconium is an interesting nucleus where a single $\beta$-decay is possible~\cite{AME20}, and it is a case study for shape coexistence in atomic nuclei~\cite{Pri22}.  The low isotopic abundance of the parent nucleus of 2.80$\%$  explains a handful of measurements. The results for 2$\beta^{-}(2\nu)$ transition in $^{96}$Zr~\cite{1993KA12,1999AR25,2001WI17,2010AR07,2018MA51} are graphically depicted in Fig.\ref{fig:96Zr}. Both direct and geochemical results are consistent in zirconium.

\subsubsection{$^{100}$Mo}

\begin{figure}[ht]
 \centering
 \includegraphics[width=0.8\textwidth]{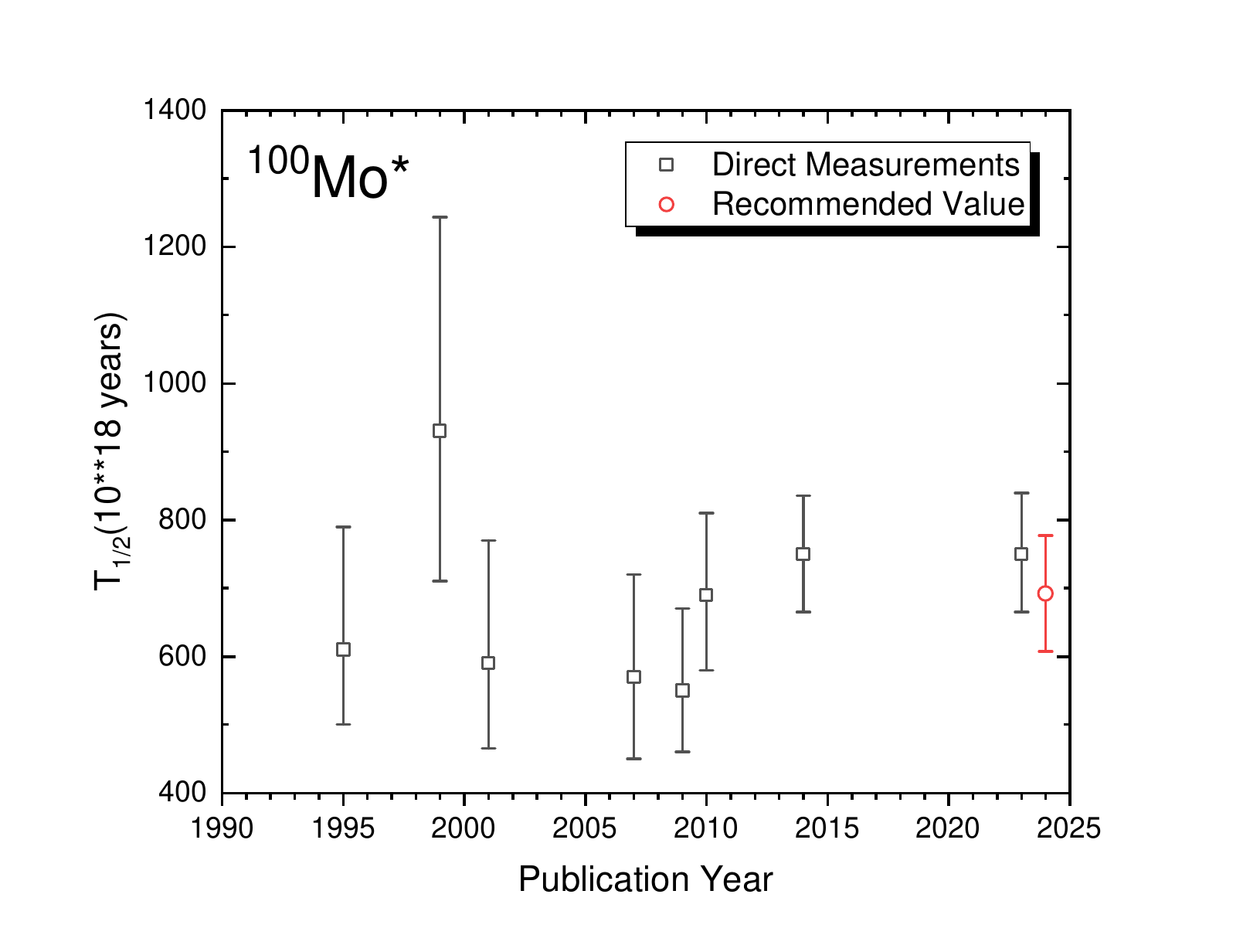}
 \includegraphics[width=0.8\textwidth]{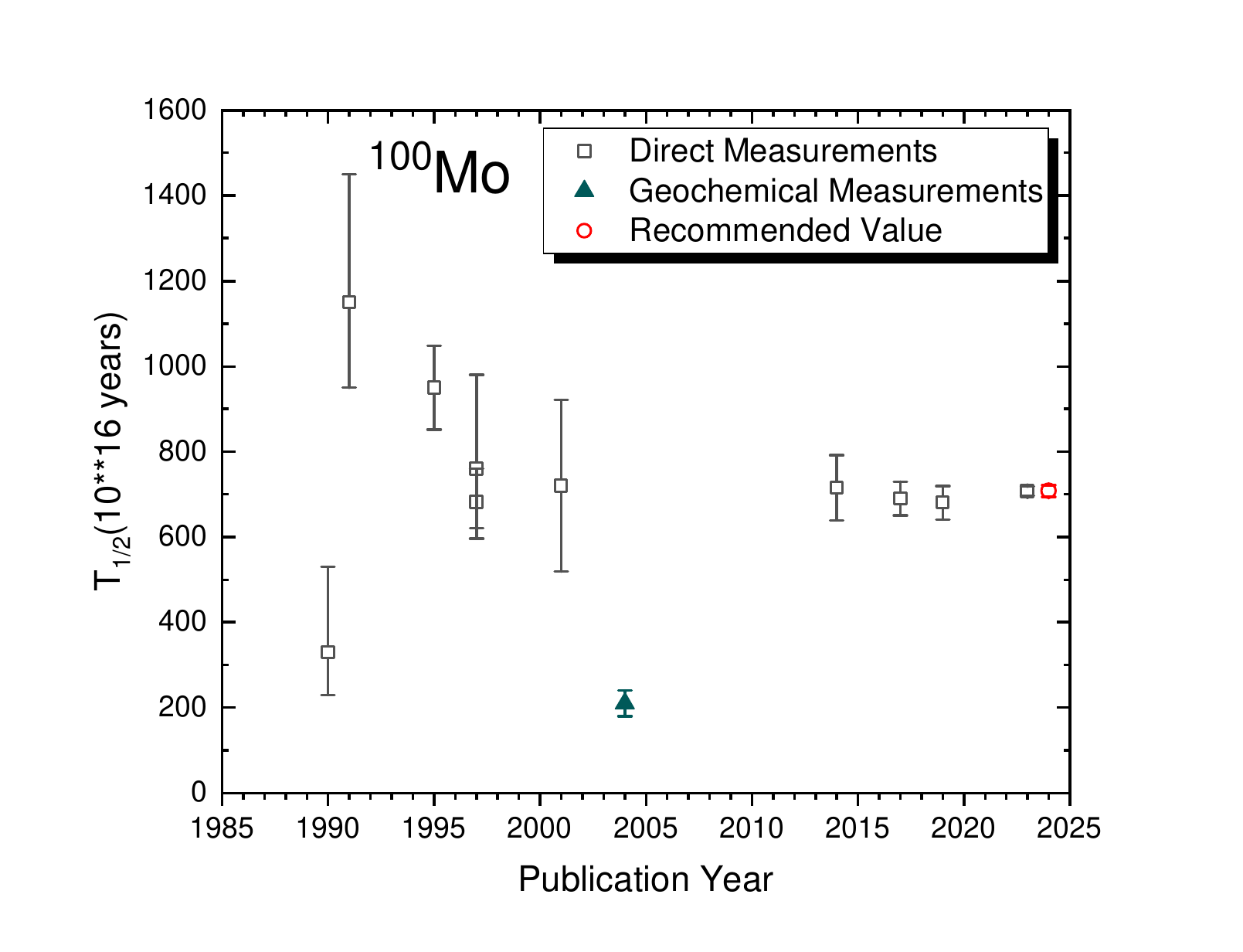}
\caption{The $^{100}$Mo ground to excited state~\cite{1995BA29,1999BB19,2001DE17,2007AR02,2006HO17,2009KI04,2010BE34,2014AR08,2023AU01} and to  the ground state~\cite{1990VA10,1991EJ02,1995DA37,1997AL02,1991EL04,1997DE40,2001AS06,2004HI19,2014CA46,2017AR18,2005AR27,2019AR04,2023AU05} transitions are shown 
 in the top and bottom panels, respectively.}
\label{fig:100Mo}
\end{figure}

Molybdenum is a well-studied nucleus where two-neutrino mode was detected for ground-to-ground and ground-to-excited state transitions. The first indication of the two-neutrino mode was obtained with plastic scintillators~\cite{1990VA10}, verified with TPC  measurements~\cite{1991EL04,1997DE40}, and extensively explored by the NEMO (Neutrino Experiments in Molybdenum) collaboration~\cite{2005AR27,1995DA37,2019AR04}. Gamma rays from ground-to-excited states were measured with ultralow-background germanium detectors. The results for 2$\beta^-$(0$^{+}_{1}$$\rightarrow$0$^{+}_{1}$)~\cite{1990VA10,1991EJ02,1995DA37,1997AL02,1991EL04,1997DE40,2001AS06,2004HI19,2014CA46,2017AR18,2005AR27,2019AR04,2020AR09,2023AU05}  and  2$\beta^-$(0$^{+}_{1}$$\rightarrow$0$^{+}_{2}$)~\cite{1995BA29,1999BB19,2001DE17,2007AR02,2006HO17,2009KI04,2010BE34,2014AR08,2023AU01}  transitions in $^{100}$Mo are graphically depicted in Fig.\ref{fig:100Mo}.

\subsubsection{$^{116}$Cd}

\begin{figure}
\centering
\includegraphics[width=0.8\textwidth]{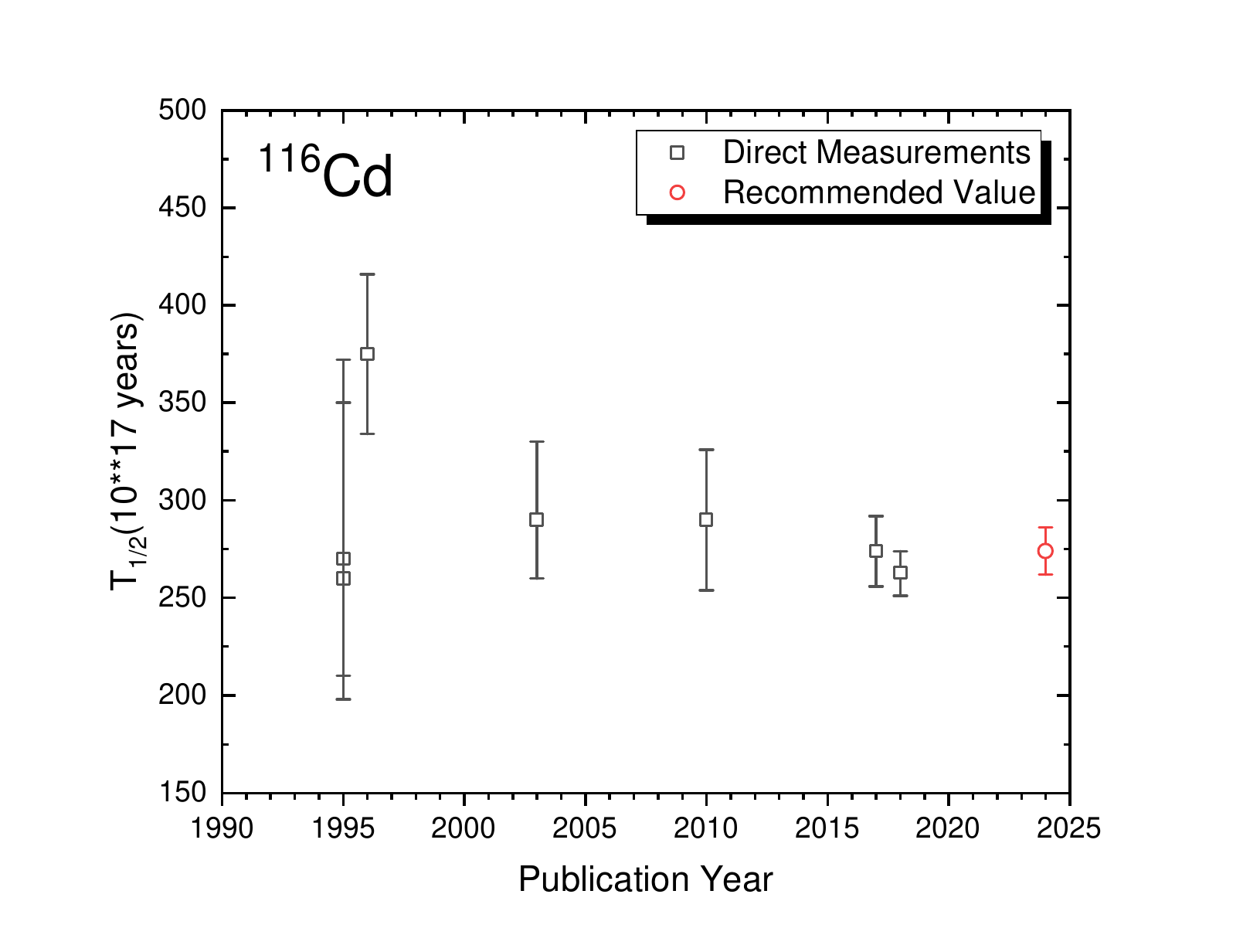}
\caption{$^{116}$Cd experimental half-lives~\cite{1995EJ01,1995DA09,1995AR26,1996AR36,ASBar10,2000DA27,2003DA09,2003DA24,2017AR01,2018BA44}.}
\label{fig:116Cd}
\end{figure}

Discovery of 2$\beta^{-}(2\nu)$ decay of $^{116}$Cd was made with the ELEGANTS V setup based on NaI(Tl) detectors, drift chambers, and thin $^{116}$Cd and $^{nat}$Cd foils~\cite{1995EJ01}. Capitalizing on the fact that cadmium is a component of scintillation materials, CdWO$_{4}$ detectors were used by the Kyiv group~\cite{1995DA09} to observe double-beta decay in $^{116}$Cd. Over the years multiple techniques were employed for the investigation of this isotope. The results for 2$\beta^{-}(2\nu)$ transition in $^{116}$Cd~\cite{1995EJ01,1995DA09,1995AR26,1996AR36,ASBar10,2000DA27,2003DA09,2003DA24,2017AR01,2018BA44} are graphically depicted in Fig.\ref{fig:116Cd}.

\subsubsection{$^{128}$Te}
\begin{figure}[ht]
 \centering
\includegraphics[width=0.8\textwidth]{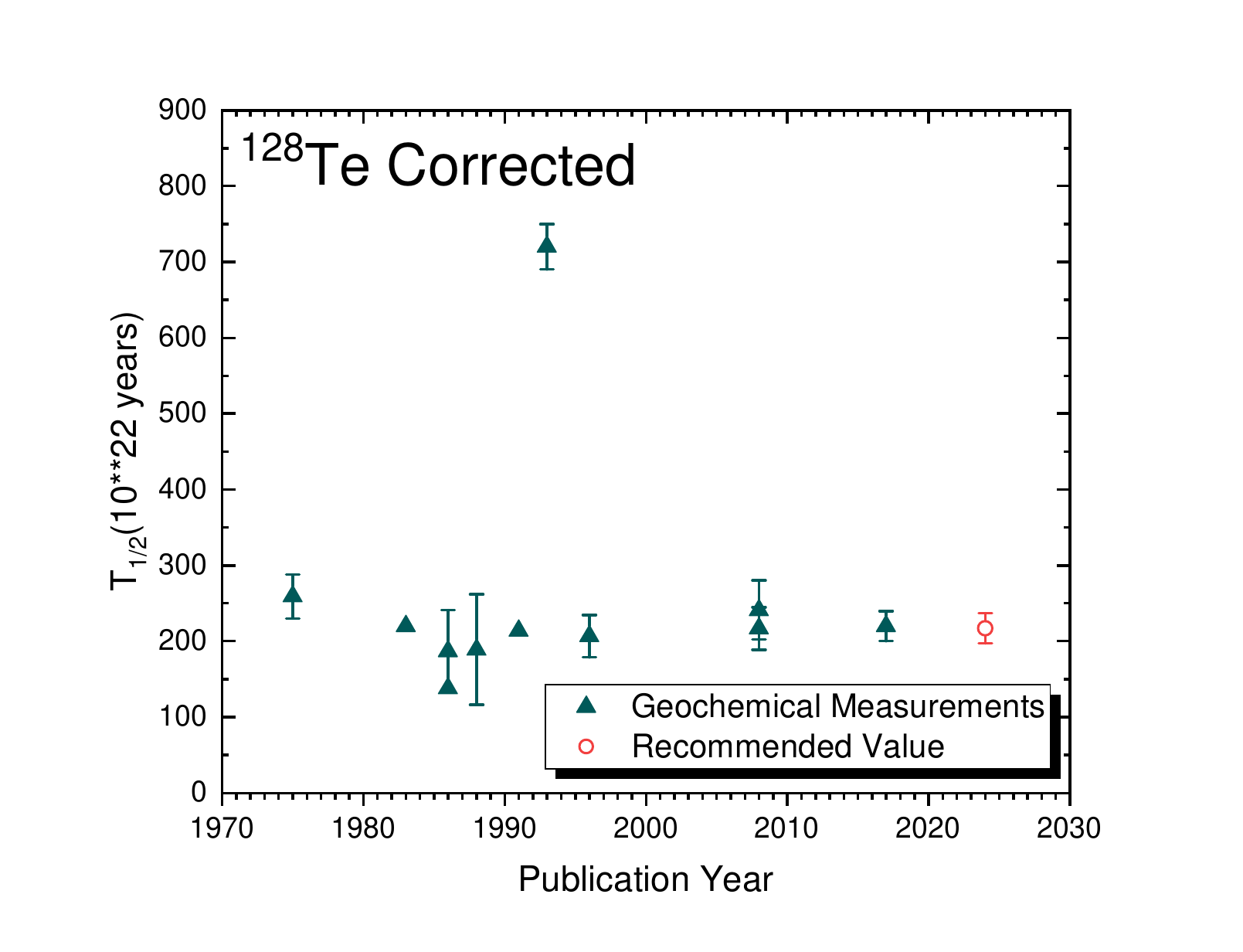}
\includegraphics[width=0.8\textwidth]{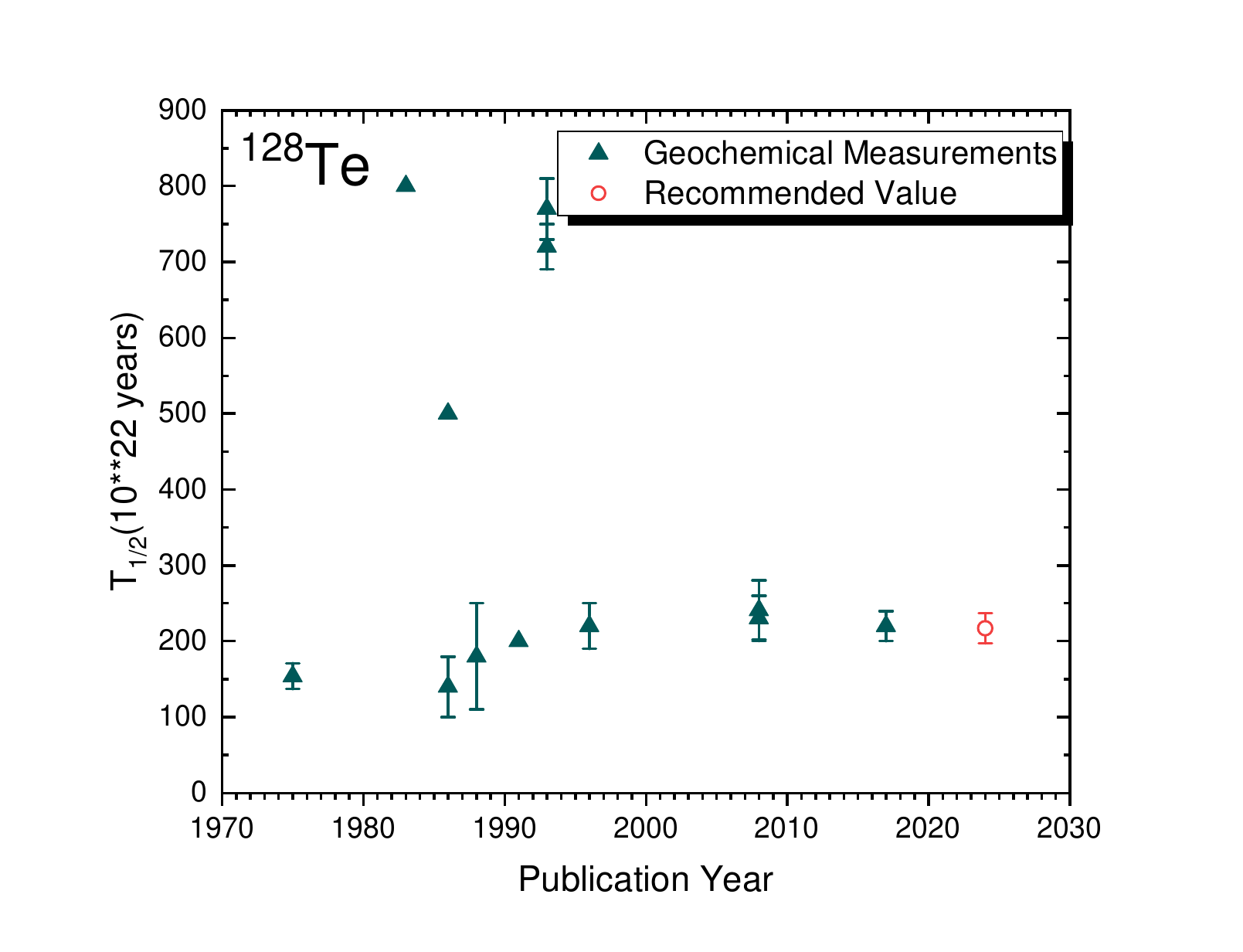}
\caption{$^{128}$Te for published~\cite{1975HE04,1983KI02,1983KI03,1986MAYT,1986KIZQ,1988LI12,1991MA64,1992BE30,1993BE04,1993Da19,1996TA04,2008ME10,2008TH07,2017MEZS} and corrected transitions are shown 
 in the bottom and top panels, respectively.}
\label{fig:128Te}
\end{figure}

The decay energy for $^{128}$Te is less than 1 MeV, and only geochemical observations are available. The results for published~\cite{1975HE04,1983KI02,1983KI03,1986MAYT,1986KIZQ,1988LI12,1991MA64,1992BE30,1993BE04,1993Da19,1996TA04,2008ME10,2008TH07} and corrected  transitions in $^{128}$Te are graphically depicted in Fig.\ref{fig:128Te}.

The $^{128}$Te half-lives were not explicitly measured, and rather deduced from the $^{130}$Te values using the $^{128}$Te/$^{130}$Te half-life ratio. The ratio was not precisely known in the past, and the earlier findings for  $^{128}$Te half-lives were affected by this issue.  In light of this disclosure, it is reasonable to follow the standard nuclear data evaluation practices and correct the shifting back and forth half-lives~\cite{Sea21,Zer18,Pri15}.  The corrected T$_{1/2}$ based on the  $^{130}$Te/$^{128}$Te half-life ratio of (3.74$\pm$0.10)$\times$10$^{-4}$~\cite{1993Da19} are shown in Table~\ref{table:128Te}. The presently selected ratio and the sheer volume of directly measured $^{130}$Te values suggest the preference for the lower geochemical T$_{1/2}$ values (geologically young samples) over the higher values for geologically old minerals~\cite{2008ME10}. The current finding is in agreement with nuclear systematics~\cite{2023PRP}.

The present-day data provide a clear preference for the lower half-life values. At the same time, $^{130}$Te(2$\beta^{-}$) direct measurements are calorimetric or 2$\beta$Fit as defined in this work. The evolution of $^{76}$Ge half-lives shows that calorimetric values strongly depend on radioactive background and have tendencies to rise. If the future $^{130}$Te data trends would follow $^{76}$Ge then it is possible that direct measurements could be consistent with the older geological materials. The latter scenario is unlikely today, however, it cannot be excluded in the future and more advanced measurements are needed.

The current analysis of double-beta decay data shows that $^{128}$Te is the most long-lived parent nucleus. Furthermore, the $^{128}$Te(2$\beta^{-}$) process has the largest half-life for a nuclear decay in nature, and its precise determination is very important.
\begin{landscape}
\begin{table}[ht] 
\caption{Re-analysis of the $^{128}$Te 2$\beta^-$(0$^{+}_{1}$$\rightarrow$0$^{+}_{1}$) published half-lives. The T$_{1/2}$ were corrected assuming the  $^{130}$Te/$^{128}$Te half-life ratio of (3.74$\pm$0.10)$\times$10$^{-4}$~\cite{1993Da19} (the (3.52$\pm$0.11)$\times$10$^{-4}$ ratio was corrected due to cosmic-ray produced Xe). $^{\star}$  character marks the cases where corrections were not needed. } 
\centering      
\begin{tabular}{c|c|c|c}  
\hline\hline
Published T$_{1/2}$ & Publication T$_{1/2}$ Ratio & Reference  & Corrected T$_{1/2}$ \\
\hline
154$^{+17}_{-17}\times10^{22}$ &  $^{128}$Te/$^{130}$Te=(1.59$\pm$0.05)$\times$10$^{+3}$  &  \cite{1975HE04} & 259$^{+29}_{-29}\times10^{22}$   \\      
 $>$800$^{+0}_{-0}\times10^{22}$ &  $^{128}$Te/$^{130}$Te=(0.971$^{+885}_{-971}$)$\times$10$^{+4}$ &  \cite{1983KI02,1983KI03} &  $>$220$^{+0}_{-0}\times10^{22}$  \\ 
140$^{+40}_{-40}\times10^{22}$  & $^{128}$Te/$^{130}$Te=2$\times$10$^{+3}$  &     \cite{1986MAYT} & 187$^{+54}_{-54}\times10^{22}$    \\    
$>$500$^{+0}_{-0}\times10^{22}$ &   $^{128}$Te/$^{130}$Te=(0.971$^{+885}_{-971}$)$\times$10$^{+4}$ &     \cite{1986KIZQ} &  $>$138$^{+0}_{-0}\times10^{22}$  \\ 
180$^{+70}_{-70}\times10^{22}$  &  $^{128}$Te/$^{130}$Te=2550$\pm$1300,  2540$\pm$680 &   \cite{1988LI12}  &  189$^{+73}_{-73}\times10^{22}$   \\ 
 200$^{+0}_{-0}\times10^{22}$  &  $^{130}$Te/$^{128}$Te=4$\times$10$^{-4}$ &     \cite{1991MA64}  &  214$^{+0}_{-0}\times10^{22}$   \\   
720$^{+30}_{-30}\times10^{22}$  & $^{130}$Te/$^{128}$Te=(3.74$\pm$0.10)$\times$10$^{-4}$ &     \cite{1992BE30,1993BE04,1993Da19}  &   720$^{+30}_{-30}\times10^{22}$$^{\star}$   \\  
220$^{+30}_{-30}\times10^{22}$  & $^{130}$Te/$^{128}$Te=(3.52$\pm$0.11)$\times$10$^{-4}$  &     \cite{1996TA04}  &  207$^{+28}_{-28}\times10^{22}$   \\       
241$^{+39}_{-39}\times10^{22}$  & $^{130}$Te/$^{128}$Te=(3.74$\pm$0.10)$\times$10$^{-4}$ &     \cite{2008ME10}  &  241$^{+39}_{-39}\times10^{22}$$^{\star}$  \\  
230$^{+30}_{-30}\times10^{22}$  & $^{130}$Te/$^{128}$Te=(3.52$\pm$0.11)$\times$10$^{-4}$  &     \cite{2008TH07}  &  217$^{+28}_{-28}\times10^{22}$  \\     
220$^{+20}_{-20}\times10^{22}$  & $^{130}$Te/$^{128}$Te=(3.74$\pm$0.10)$\times$10$^{-4}$ &     \cite{2017MEZS}  &  220$^{+20}_{-20}\times10^{22}$$^{\star}$  \\  
\hline\hline     
\end{tabular} 
\label{table:128Te}  
\end{table} 
\end{landscape}

\subsubsection{$^{130}$Te}
\begin{figure}
\centering
\includegraphics[width=0.8\textwidth]{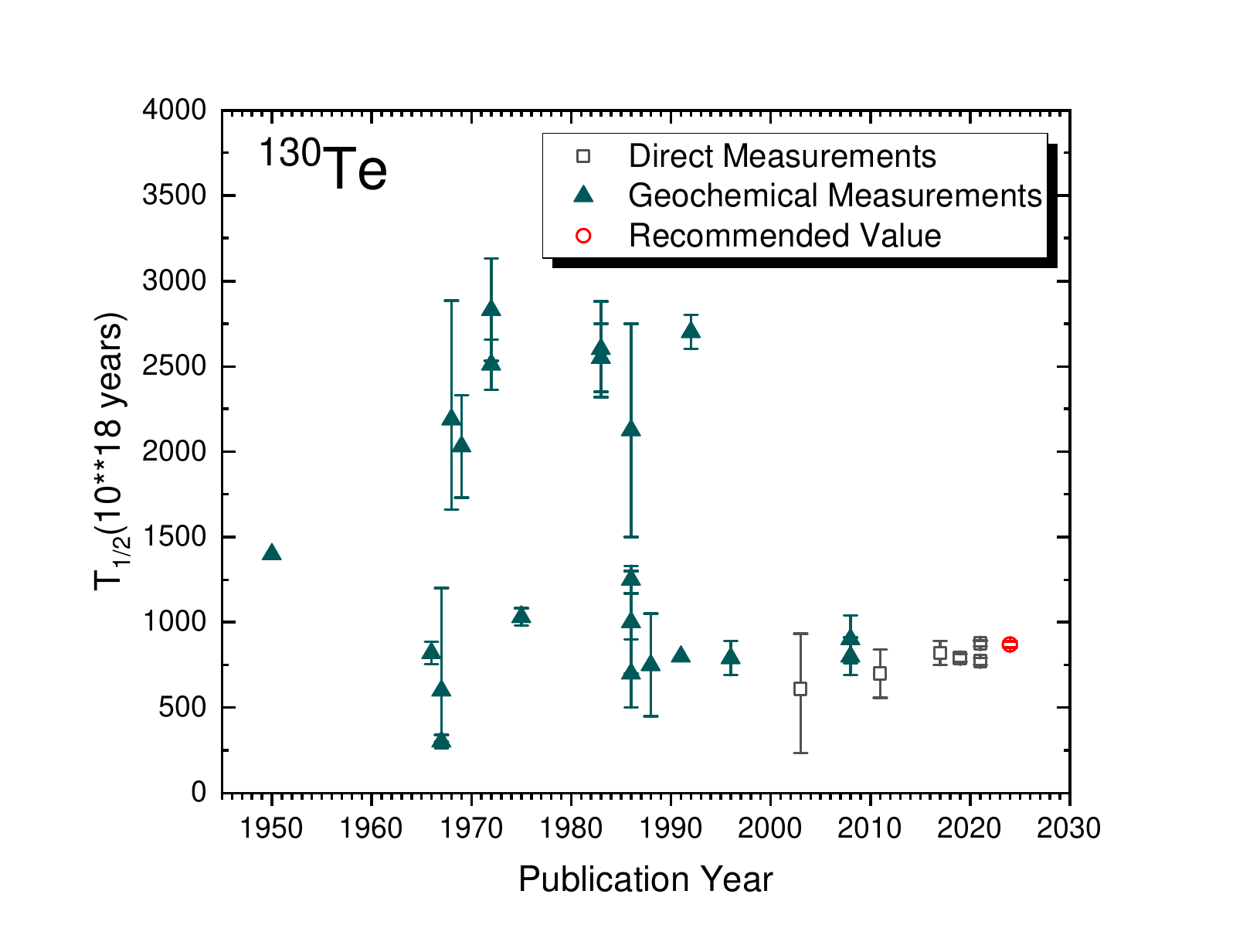}
\caption{$^{130}$Te geochemical~\cite{1950IN03,1966TA02,1967KI04,1967GE12,1968KI02,1969AL22,1972SR03,1972SR06,1975HE04,1983KI02,1983KI03,1983KIZV,1986MAYT,1986KIZQ,1986RI02,1986LI10,1988LI11,1991MA64,1992BE30,1993BE04,1996TA04,2008ME10,2008TH07} and directly measured half-lives~\cite{2003AR02,2011AR09,2017AL46,2019CA21,2021AD18,2021AD18E}.}
\label{fig:130Te}
\end{figure}

The results for 2$\beta^{-}(2\nu)$ transition in $^{130}$Te~\cite{1950IN03,1966TA02,1967KI04,1967GE12,1968KI02,1969AL22,1972SR03,1972SR06,1975HE04,1983KI02,1983KI03,1983KIZV,1986MAYT,1986KIZQ,1986RI02,1986LI10,1988LI11,1991MA64,1992BE30,1993BE04,1996TA04,2008ME10,2008TH07,2003AR02,2011AR09,2017AL46,2021AD18,2021AD18E} are graphically depicted in Fig.\ref{fig:130Te}, and large discrepancies are clearly visible.
This isotope of tellurium was first measured geochemically and later directly. 

The geological half-lives for Te specimens of known age fall into two distinct groups: $\sim$2.5$\times$10$^{21}$ and $\sim$8$\times$10$^{20}$ yr for geologically old ($>$1 Gyr) and young ($\sim$100 Myr) Te samples~\cite{2008ME10}, respectively. These discrepancies between geochemical experiments are due to many wild card parameters such as the exact age of the specimen, its geological history, etc. which lead to greater uncertainties and less reliable findings in the older samples. There are two possible scenarios here, first, we can process all data sets simultaneously~\cite{Pri14} or, second, separate them into two batches and analyze them individually. 

Fortunately, in $^{130}$Te case, we have the third batch based on direct calorimetric measurements that supply an independent benchmark. The young samples' results are consistent with the direct measurements while the older samples are in disagreement. Here, as in $^{82}$Se case, we select direct measurements over geochemical to deduct evaluated half-lives.

\subsubsection{$^{124}$Xe}
\begin{figure}
\centering
\includegraphics[width=0.8\textwidth]{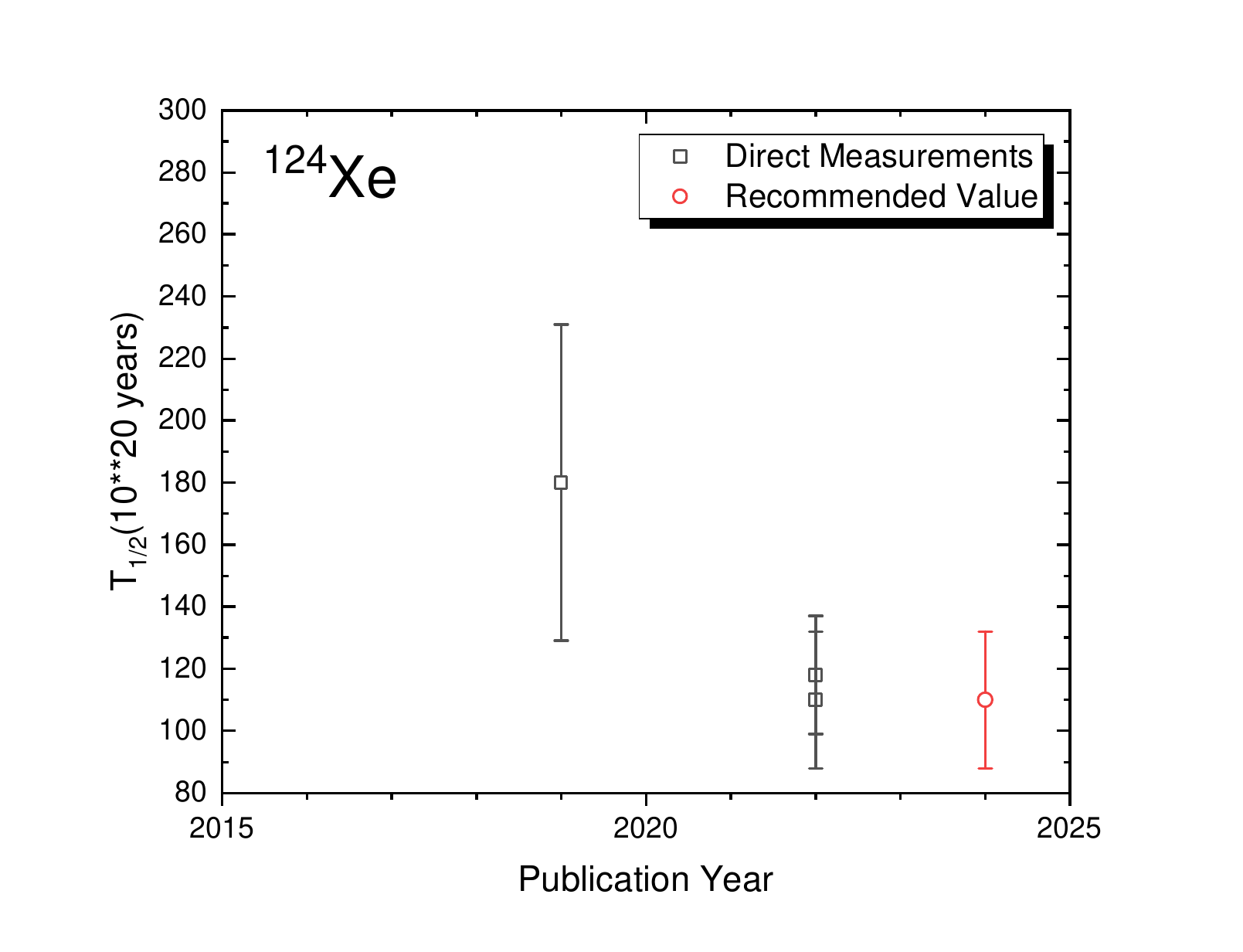}
\caption{$^{124}$Xe experimental half-lives~\cite{2019AP03,2022AP04,2022AP01}.}
\label{fig:124Xe}
\end{figure}

Xenon has the largest Q-value for 2$\epsilon$-capture among even-even stable nuclei, and it is used in gaseous (or liquid) detectors. Two-neutrino double electron capture in xenon was detected with the XENON1T dark-matter detector~\cite{2019AP03}.  The results for 2$\epsilon(2\nu)$ transition in $^{124}$Xe using XENON1T and XENONnT detectors~\cite{2019AP03,2022AP04,2022AP01} are graphically depicted in Fig.\ref{fig:124Xe}. Small isotopic abundance explains the limited number of measurements for this isotope.

\subsubsection{$^{136}$Xe}

\begin{figure}
\centering
\includegraphics[width=0.8\textwidth]{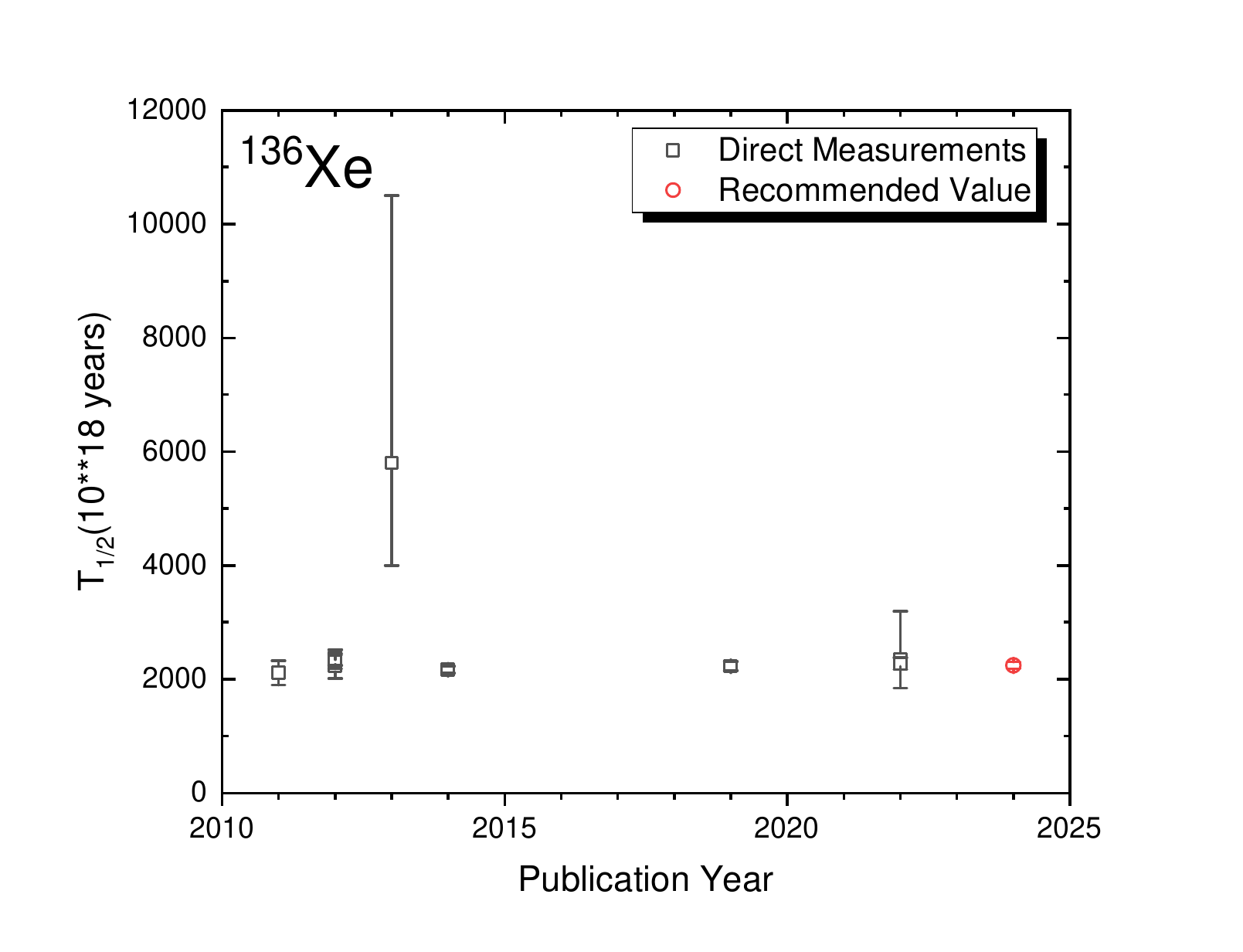}
\caption{$^{136}$Xe experimental half-lives~\cite{2002BE74,2011AC03,2012AU03,2012GA17,2012GA32,2013GA41,2014AL03,2016GA30,2019GA11,2022NO06,2022SIZZ}.}
\label{fig:136Xe}
\end{figure}

Xenon is another nucleus where the detector's active volume consists of double-beta decay parent nuclei. The energy resolution of Xe calorimeters is lower than in Ge-detectors while Q-value is higher. Due to xenon production issues substantial amounts of radioactive $^{85}$Kr is present in the gas.  The krypton single $\beta$-decay introduces an additional radioactive background below 700 keV in xenon detectors that impacts the 2$\nu$ spectrum.  The results for 2$\beta^{-}(2\nu)$ transition in $^{136}$Xe~\cite{2002BE74,2011AC03,2012AU03,2012GA17,2012GA32,2013GA41,2014AL03,2016GA30,2019GA11,2022NO06,2022SIZZ} are graphically depicted in Fig.\ref{fig:136Xe}. Most of the results are consistent except for a single outlier with large experimental errors~\cite{2013GA41}.

\subsubsection{$^{130}$Ba}

\begin{figure}
\centering
\includegraphics[width=0.8\textwidth]{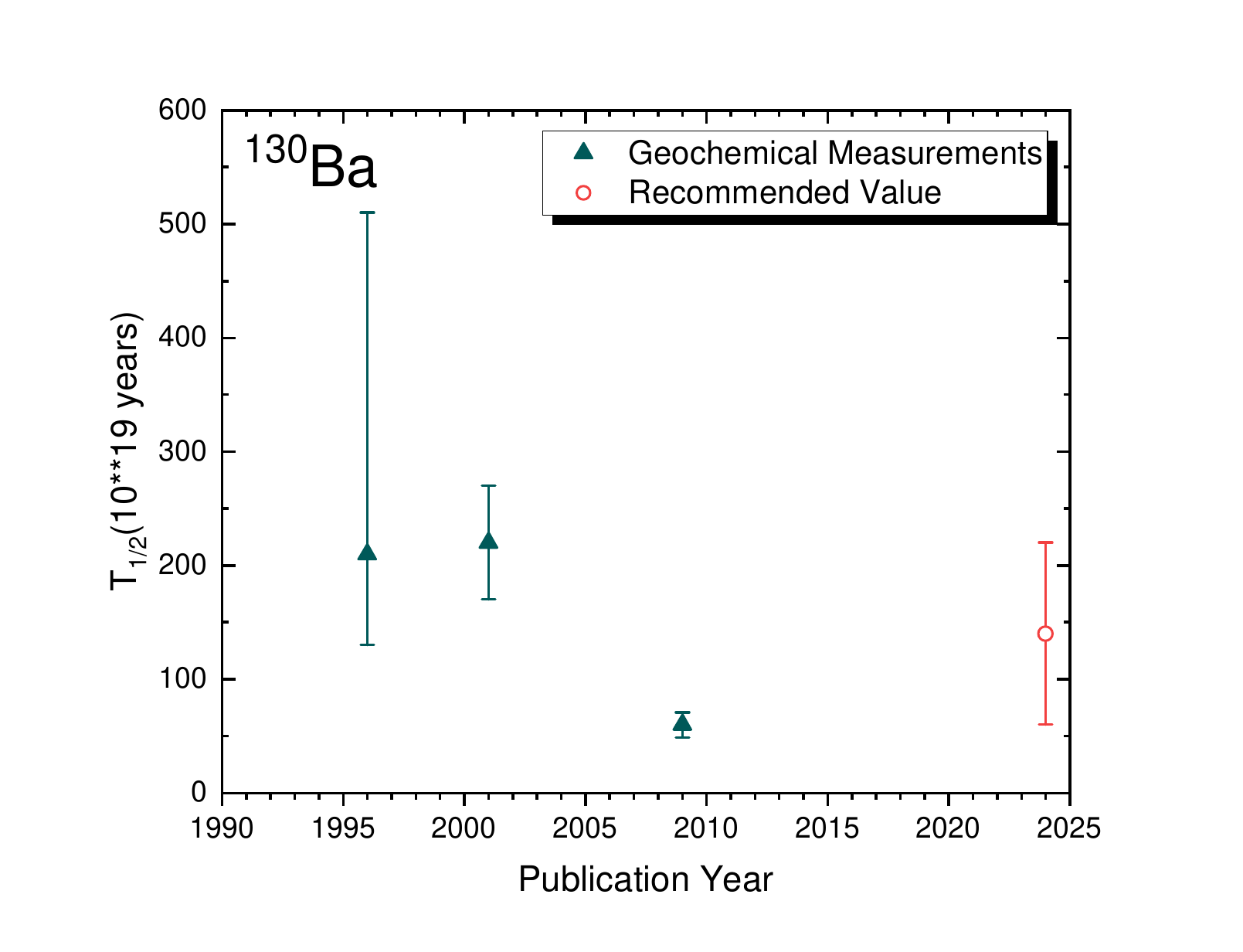}
\caption{$^{130}$Ba experimental half-lives~\cite{1996BA24,2001ME22,2009PU05}.}
\label{fig:130Ba}
\end{figure}

The barium nucleus has a very low isotopic abundance of 0.11$\%$, and only geochemical measurements are presently available. 
The results for 2$\epsilon(2\nu)$ transition in $^{130}$Ba~\cite{1996BA24,2001ME22,2009PU05} are graphically depicted in Fig.\ref{fig:130Ba}.

\subsubsection{$^{150}$Nd}
\begin{figure}[ht]
 \centering
\includegraphics[width=0.8\textwidth]{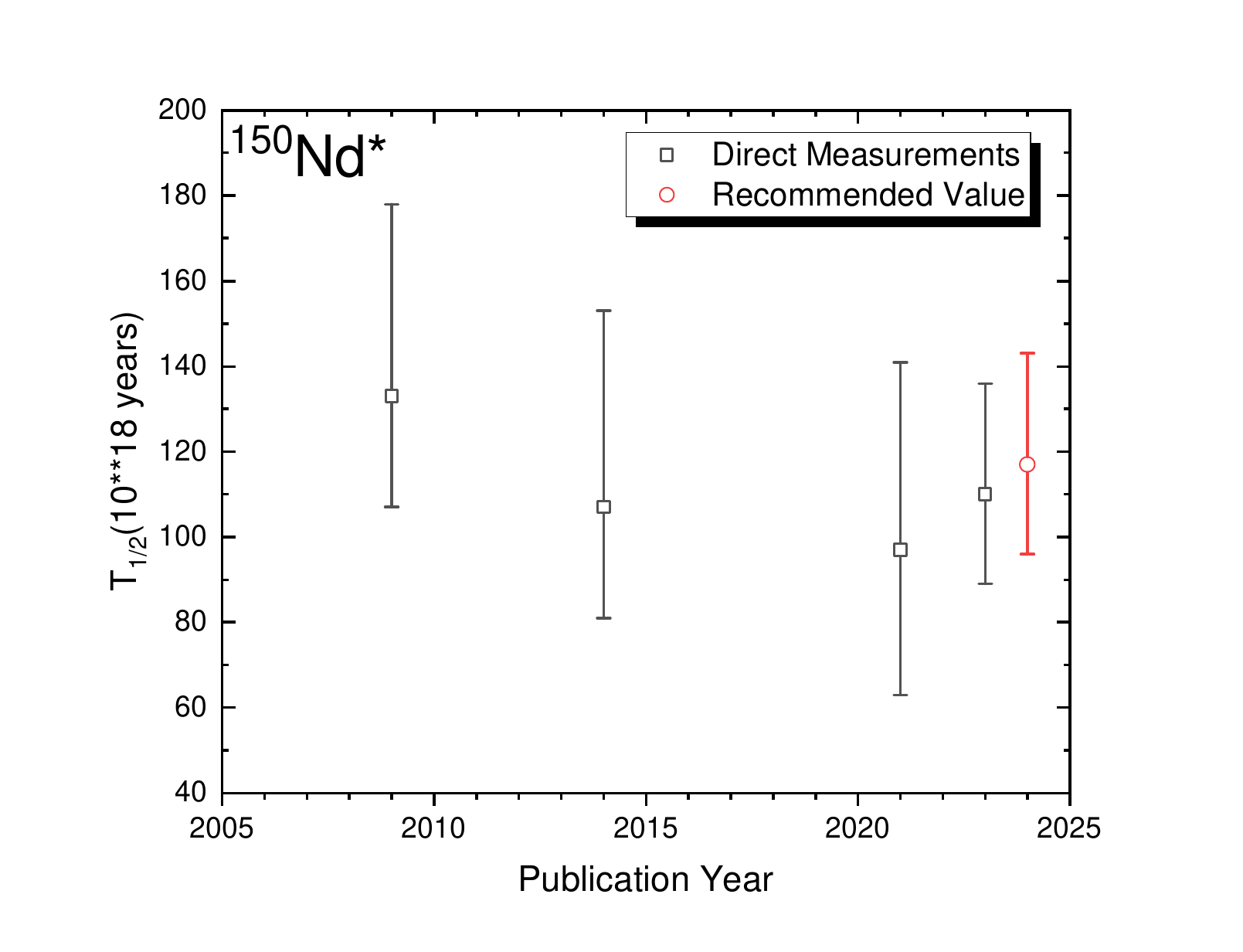}
\includegraphics[width=0.8\textwidth]{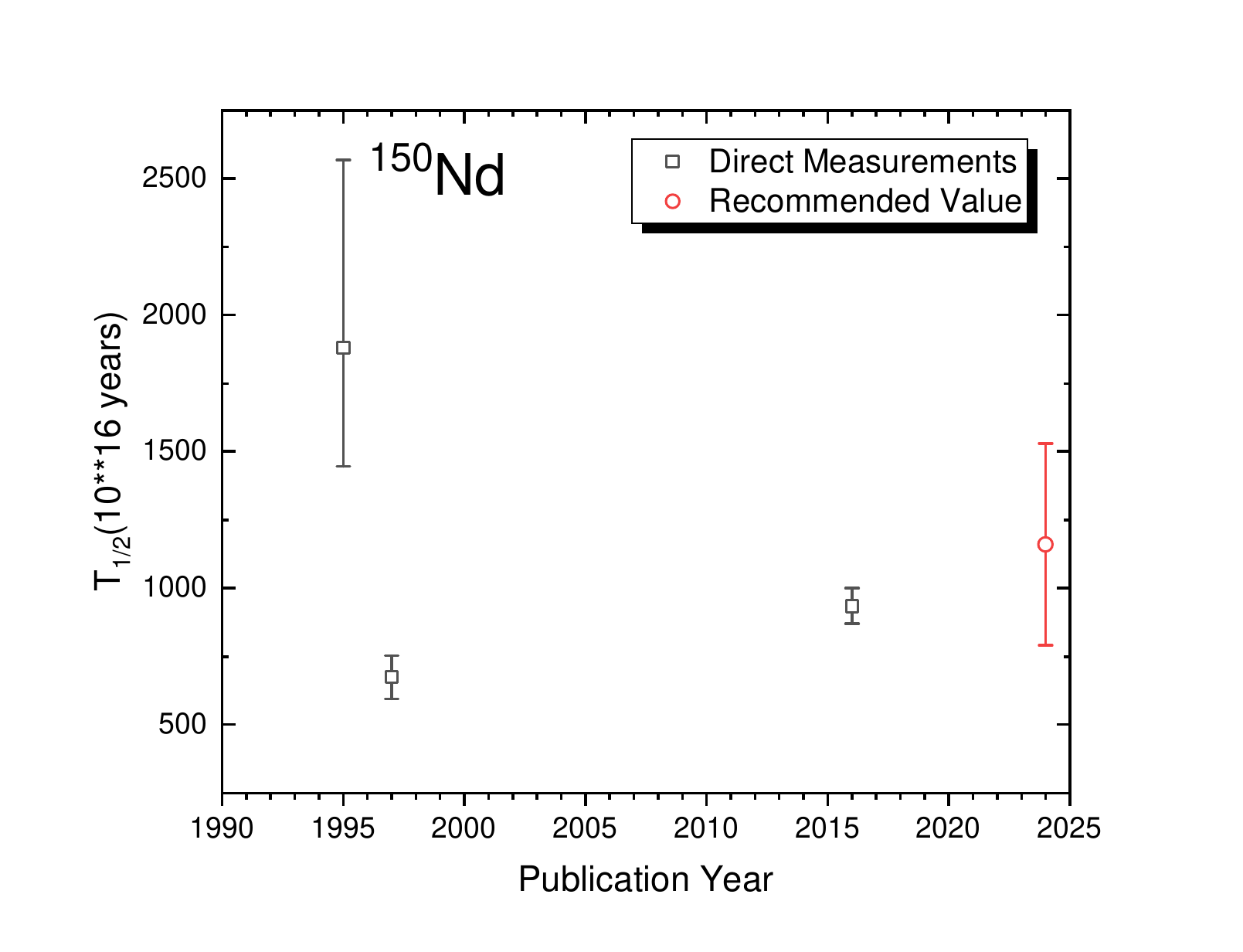}
\caption{The $^{150}$Nd ground to the excited state~\cite{2004BA10,2009BA21,2014KI13,2021PO16,2023AG06} transitions and the ground state~\cite{1995AR08,1997DE40,2016AR12} transitions  are shown 
 in the top and bottom panels, respectively.}
\label{fig:150Nd}
\end{figure}

Neodymium is the second nucleus where two-neutrino mode was detected for ground-to-ground and ground-to-excited state transitions. It has the second-highest double-beta transition Q-value among stable even-even nuclei, and $^{150}$Nd$\rightarrow ^{150}$Sm transition may be hindered by the intermediate nucleus $^{150}$Pm ground-state negative parity.  The two-neutrino transition in neodymium was first discovered with ITEP TPC~\cite{1995AR08}, verified with the UC-Irvine TPC~\cite{1997DE40}, and finalized in NEMO-3 experiment~\cite{2009AR10,2016AR12}.  Transitions from the ground-to-excited states were measured with low-background HPGe  $\gamma$-ray detectors. 
The results for 2$\beta^-$(0$^{+}_{1}$$\rightarrow$0$^{+}_{1}$)~\cite{1995AR08,1997DE40,2016AR12}  and  2$\beta^-$(0$^{+}_{1}$$\rightarrow$0$^{+}_{2}$)~\cite{2004BA10,2009BA21,2014KI13,2021PO16,2023AG06}  transitions in $^{150}$Nd are graphically depicted in Fig.\ref{fig:150Nd}.

\subsubsection{$^{238}$U}

\begin{figure}
\centering
\includegraphics[width=0.8\textwidth]{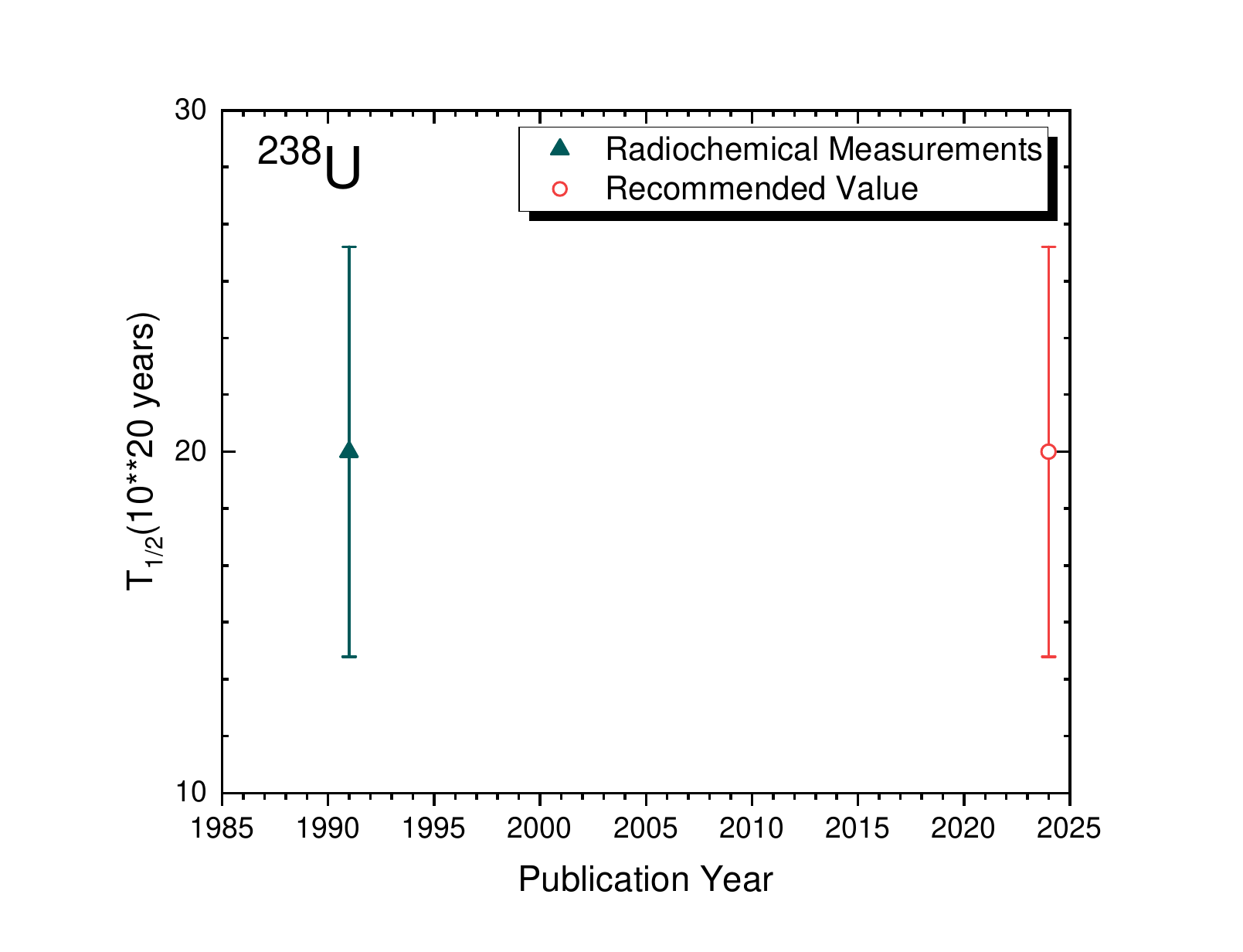}
\caption{$^{238}$U radiochemical half-life~\cite{1991TU02}.}
\label{fig:238U}
\end{figure}

The half-life for the decay of $^{238}$U to $^{238}$Pu has been measured by chemically isolating and measuring from the resultant alpha particles the amount of plutonium that had accumulated in 35 yr from 8.4 kg of purified uranyl nitrate. It is a radiochemical observation of daughter nuclei under controlled conditions (clean geochemical measurement). The results for 2$\beta^{-}(2\nu)$ transition in $^{238}$U~\cite{1991TU02} are graphically depicted in Fig.\ref{fig:238U}.

\section{Data Evaluation and Recommended Values}
The recommended $2 \beta$(2$\nu$)-decay half-lives were deduced using the standard USNDP procedures~\cite{ensdf,Dim20,NSDD}. The present work evaluation policies and evaluated (recommended) half-lives are given below.

\subsection{Evaluation Policies}
\begin{itemize}
\item Compile final values of double-beta decay observations
\item Analyze the final results in detail
\item Round uncertainties to two (rarely three) significant digits
\item Accept asymmetric uncertainties, if necessary
\item Assign 10$\%$ uncertainties when no uncertainties are given by authors
\item Accept evaluated uncertainties of higher or equal of the lowest experimental values
\item Direct communication with authors in discrepant cases, if possible
\item Give preference to direct over geochemical measurements if both types are available
\item Deduce recommended half-lives using the visual averaging library~\cite{VisLib} for selected data sets
\item Accept weighted value if reduced $\chi^2$ is reasonable, for discrepant data adopt unweighted average values
\end{itemize}

\subsection{Evaluated Half-Lives}
The evaluated half-lives are given in Table~\ref{Recom}. We used the latest release of the visual averaging library~\cite{VisLib} and presently available data. The visual averaging library program includes unweighted and weighted averages as well as the limitation of relative statistical weights~\cite{LWM}, normalized residual~\cite{NRM}, Rajeval technique~\cite{RT}, the Expected Value~\cite{EVM}, bootstrap and Mandel-Paule~\cite{BMP,Ruk2009} statistical methods to calculate averages of experimental data with uncertainties. 

In our evaluation, we generally adopted
the weighted average, using the unweighted average in several discrepant cases. Routinely, the unweighted average is used when reduced $\chi^{2}>1$. However, due to the discrepant nature of double-beta decay data sets~\cite{Pri15}, we accepted reduced $\chi^{2}< 2$ for the data fits. In several cases, the Rajeval technique produces much lower values of $\chi^2/(N-1)$ than other methods. This issue is particularly clear for $^{130}$Ba data sets when several methods produce identical results with different $\chi^2/(N-1)$ values. Therefore, we could not agree with the Rajeval treatment of uncertainties and $\chi^2/(N-1)$ values and rejected these results. Finally, we adopted evaluated values uncertainties consistent (higher or equal) with experimental uncertainties. 

\begin{landscape}
\begin{longtable}[c]{l c| c| c| c| c} 
\caption{Evaluation of experimental half-lives for $2 \beta$(2$\nu$)-decay. Ground-to-ground state transitions are assumed unless daughter nucleus  $\gamma$-rays were measured.} \\
\hline\hline
   \textbf{Nuclide}  &   \textbf{Decay} &   \textbf{Method} &  \textbf{T$_{1/2}^{2 \nu}$(y)} & $\chi^2/(N-1)$ & \textbf{Evaluated T$_{1/2}^{2 \nu}$(y)}  \\
 \hline 
\endfirsthead
\hline
   \textbf{Nuclide}  &   \textbf{Decay} &   \textbf{Method} &  \textbf{T$_{1/2}^{2 \nu}$(y)} & $\chi^2/(N-1)$ & \textbf{Evaluated T$_{1/2}^{2 \nu}$(y)}  \\
\hline 
\endhead
\hline
\multicolumn{5}{r}{\textit{Continued on next page ...}} \\
\endfoot
\endlastfoot
$^{48}$Ca & 2$\beta^-$(0$^{+}_{1}$$\rightarrow$0$^{+}_{1}$) & Unweighted Average  & 4.970(720)$\times$10$^{19}$   &   & \\      
                   &                                                            & Weighted Average  & 5.960$^{+1020}_{-960}$$\times$10$^{19}$  & 0.40  &  5.96$^{+139}_{-108}$$\times$10$^{19}$ \\ 
                   &                                                            & Limitation of Statistical Weights  & 5.960$^{+1020}_{-960}$$\times$10$^{19}$  & 0.40  &   \\ 
                   &                                                            & Normalized Residuals  & 5.960$^{+1020}_{-960}$$\times$10$^{19}$  & 0.40  &   \\   
                   &                                                            & Rajeval Technique  & 5.960$^{+1020}_{-960}$$\times$10$^{19}$  & 0.40  &   \\ 
                   &                                                            & Expected Value  & 5.200$^{+1370}_{-780}$$\times$10$^{19}$  &   &   \\ 
                   &                                                            & Bootstrap  & 5.800(1400)$\times$10$^{19}$  & 0.41  &   \\ 
                   &                                                            & Mandel-Paule  & 5.630$^{+1020}_{-960}$$\times$10$^{19}$  & 0.46  &   \\  
\hline  
$^{76}$Ge & 2$\beta^-$(0$^{+}_{1}$$\rightarrow$0$^{+}_{1}$) & Unweighted Average  & 1.745(89)$\times$10$^{21}$   &   &  1.745(89)$\times$10$^{21}$  \\      
                   &                                                            & Weighted Average  & 1.914(70)$\times$10$^{21}$  &  4.87  &  \\ 
                   &                                                            & Limitation of Statistical Weights  & 1.888(74)$\times$10$^{21}$  & 4.32  &   \\ 
                   &                                                            & Normalized Residuals  & 1.882(72)$\times$10$^{21}$  & 2.66  &   \\   
                  &                                                            & Rajeval Technique  & 1.815$^{+59}_{-66}$$\times$10$^{21}$  & 1.47  &    \\ 
                   &                                                            & Expected Value  & 1.790(190)$\times$10$^{21}$  &   &   \\ 
                   &                                                            & Bootstrap  & 1.770(120)$\times$10$^{21}$    & 6.61  &   \\ 
                   &                                                            & Mandel-Paule  & 1.780(180)$\times$10$^{21}$    & 6.17  &   \\  
\hline
$^{82}$Se & 2$\beta^-$(0$^{+}_{1}$$\rightarrow$0$^{+}_{1}$) & Unweighted Average  & 9.16(45)$\times$10$^{19}$   &   & 9.16(45)$\times$10$^{19}$  \\      
                   &                                                            & Weighted Average  & 8.73(16)$\times$10$^{19}$  & 3.47  &   \\ 
                   &                                                            & Limitation of Statistical Weights  & 8.76(22)$\times$10$^{19}$  & 3.42  &   \\ 
                   &                                                            & Normalized Residuals  & 8.76(22)$\times$10$^{19}$  & 3.42  &   \\   
                   &                                                            & Rajeval Technique  & 8.79(24)$\times$10$^{19}$  & 3.36  &   \\ 
                   &                                                            & Expected Value  & 8.74(40)$\times$10$^{19}$  &   &   \\ 
                   &                                                            & Bootstrap  & 9.09(71)$\times$10$^{19}$  & 5.23  &   \\ 
                   &                                                            & Mandel-Paule  & 8.68(22)$\times$10$^{19}$  & 3.55  &   \\  
\hline  
$^{78}$Kr &  2$\epsilon$(0$^{+}_{1}$$\rightarrow$0$^{+}_{1}$) & Single measurement  & 1.90$^{+133}_{-76}\times10^{22}$    &   &  1.90$^{+133}_{-76}\times10^{22}$  \\      
\hline  
$^{96}$Zr & 2$\beta^-$(0$^{+}_{1}$$\rightarrow$0$^{+}_{1}$) & Unweighted Average  & 2.23(34)$\times$10$^{19}$   &   &   \\      
                   &                                                            & Weighted Average  & 2.335(60)$\times$10$^{19}$  & 0.09  & 2.335(210)$\times$10$^{19}$  \\ 
                   &                                                            & Limitation of Statistical Weights  & 2.335(200)$\times$10$^{19}$   & 0.09   &  \\ 
                   &                                                            & Normalized Residuals  & 2.335(200)$\times$10$^{19}$   &  0.09 &   \\   
                   &                                                            & Rajeval Technique  & 2.335(200)$\times$10$^{19}$   & 0.09  &   \\ 
                   &                                                            & Expected Value  & 2.25$^{+34}_{-22}$$\times$10$^{19}$  &   &   \\ 
                   &                                                            & Bootstrap  & 2.37(34)$\times$10$^{19}$  &  0.12 &   \\ 
                   &                                                            & Mandel-Paule  & 2.33(20)$\times$10$^{19}$  & 0.09  &   \\  
\hline  
$^{100}$Mo & 2$\beta^-$(0$^{+}_{1}$$\rightarrow$0$^{+}_{1}$) & Unweighted Average  & 7.44(74)$\times$10$^{18}$   &   &   \\      
                   &                                                            & Weighted Average  & 7.08(14)$\times$10$^{18}$  & 1.92  & 7.08(14)$\times$10$^{18}$  \\ 
                   &                                                            & Limitation of Statistical Weights  & 7.08(14)$\times$10$^{18}$  & 1.92  &\\ 
                   &                                                            & Normalized Residuals  & 7.08(14)$\times$10$^{18}$  & 1.92  &   \\   
                   &                                                            & Rajeval Technique  & 7.07(12)$\times$10$^{18}$  &  1.33 &   \\ 
                   &                                                            & Expected Value  & 7.17(81)$\times$10$^{18}$  &    &  \\ 
                   &                                                            & Bootstrap  & 7.33(71)$\times$10$^{18}$  &  2.68  &  \\ 
                   &                                                            & Mandel-Paule  & 7.20(130)$\times$10$^{18}$  &  2.15  &  \\  
\hline  
$^{100}$Mo & 2$\beta^-$(0$^{+}_{1}$$\rightarrow$0$^{+}_{2}$) & Unweighted Average  & 6.80(52)$\times$10$^{20}$   &    &  \\      
                   &                                                            & Weighted Average  & 6.92(43)$\times$10$^{20}$   & 0.67   & 6.92(85)$\times$10$^{20}$ \\ 
                   &                                                            & Limitation of Statistical Weights  & 6.92(43)$\times$10$^{20}$  & 0.67  &   \\ 
                   &                                                            & Normalized Residuals  & 6.92(43)$\times$10$^{20}$  &  0.67  &  \\   
                   &                                                            & Rajeval Technique  & 6.92(43)$\times$10$^{20}$  & 0.67  &   \\ 
                   &                                                            & Expected Value  & 6.63(92)$\times$10$^{20}$  &   &   \\ 
                   &                                                            & Bootstrap  & 6.92(65)$\times$10$^{20}$  &    0.67 &  \\ 
                   &                                                            & Mandel-Paule  & 6.77(43)$\times$10$^{20}$  & 0.69  &   \\  
\hline  
$^{116}$Cd & 2$\beta^-$(0$^{+}_{1}$$\rightarrow$0$^{+}_{1}$) & Unweighted Average  & 2.89(19)$\times$10$^{19}$   &   &   \\      
                   &                                                            & Weighted Average  & 2.738(95)$\times$10$^{19}$  & 1.26  &  2.738(120)$\times$10$^{19}$  \\ 
                   &                                                            & Limitation of Statistical Weights  & 2.738(95)$\times$10$^{19}$  &  1.26  &  \\ 
                   &                                                            & Normalized Residuals  & 2.736(93)$\times$10$^{19}$  & 1.22  &   \\   
                   &                                                            & Rajeval Technique  & 2.711(85)$\times$10$^{19}$  & 0.63  &   \\ 
                   &                                                            & Expected Value  & 2.76(23)$\times$10$^{19}$  &    &  \\ 
                   &                                                            & Bootstrap  & 2.90(25)$\times$10$^{19}$  &  1.85 &   \\ 
                   &                                                            & Mandel-Paule  & 2.79(14)$\times$10$^{19}$  &  1.33  &  \\  
\hline   
$^{128}$Te & 2$\beta^-$(0$^{+}_{1}$$\rightarrow$0$^{+}_{1}$) & Unweighted Average  & 2.167(86)$\times$10$^{24}$     & &  2.167(200)$\times$10$^{24}$   \\      
                   &                                                            & Weighted Average  & 2.084(39)$\times$10$^{24}$  & 5.18  &   \\ 
                   &                                                            & Limitation of Statistical Weights  & 2.084(53)$\times$10$^{24}$  &  5.18   & \\ 
                   &                                                            & Normalized Residuals  & 2.095(33)$\times$10$^{24}$  & 2.97   &  \\   
                   &                                                            & Rajeval Technique  & 2.086(29)$\times$10$^{24}$  &  1.67  &  \\ 
                   &                                                            & Expected Value  & 2.09(15)$\times$10$^{24}$  &    &  \\ 
                   &                                                            & Bootstrap  & 2.119(94)$\times$10$^{24}$  & 5.75   & \\ 
                   &                                                            & Mandel-Paule  & 2.07(15)$\times$10$^{24}$  &  5.23  &  \\                                                                                                                 
\hline
$^{130}$Te & 2$\beta^-$(0$^{+}_{1}$$\rightarrow$0$^{+}_{1}$) & Unweighted Average  & 7.52(96)$\times$10$^{20}$   &    &  \\      
                   &                                                            & Weighted Average  & 8.69(17)$\times$10$^{20}$  &  0.94  & 8.69(18)$\times$10$^{20}$ \\ 
                   &                                                            & Limitation of Statistical Weights  & 8.69(17)$\times$10$^{20}$  &  0.94  &  \\ 
                   &                                                            & Normalized Residuals  & 8.69(17)$\times$10$^{20}$   & 0.94  &\\   
                   &                                                            & Rajeval Technique  & 8.45(35)$\times$10$^{20}$  &  0.74   & \\ 
                   &                                                            & Expected Value  & 8.27(82)$\times$10$^{20}$  &     & \\ 
                   &                                                            & Bootstrap  & 7.70(100)$\times$10$^{20}$  &  10.98  &  \\ 
                   &                                                            & Mandel-Paule  & 8.69(17)$\times$10$^{20}$  & 0.94    & \\     
\hline  
$^{124}$Xe &  2$\epsilon$(0$^{+}_{1}$$\rightarrow$0$^{+}_{1}$) & Single measurement  & 1.10(22)$\times10^{22}$   &    & 1.10(22)$\times10^{22}$  \\                     
\hline
$^{136}$Xe & 2$\beta^-$(0$^{+}_{1}$$\rightarrow$0$^{+}_{1}$) & Unweighted Average  & 2.860(590)$\times$10$^{21}$   &     & \\      
                   &                                                            & Weighted Average  & 2.240(46)$\times$10$^{21}$  &  1.24   & 2.240(61)$\times$10$^{21}$ \\ 
                   &                                                            & Limitation of Statistical Weights  & 2.219(41)$\times$10$^{21}$  & 0.56   &  \\ 
                   &                                                            & Normalized Residuals  & 2.240(46)$\times$10$^{21}$  & 1.24   &  \\   
                   &                                                            & Rajeval Technique  & 2.240(46)$\times$10$^{21}$  & 1.24   &  \\ 
                   &                                                            & Expected Value  & 2.260(180)$\times$10$^{21}$  &     & \\ 
                   &                                                            & Bootstrap  & 2.470(600)$\times$10$^{21}$  &  8.29  &  \\ 
                   &                                                            & Mandel-Paule  & 2.219(46)$\times$10$^{21}$  &  1.24   & \\ 
\hline
$^{130}$Ba &  2$\epsilon$(0$^{+}_{1}$$\rightarrow$0$^{+}_{1}$) & Unweighted Average  & 1.40(80)$\times$10$^{21}$   &    & 1.40(80)$\times$10$^{21}$ \\      
                   &                                                            & Weighted Average  & 6.7(34)$\times$10$^{20}$  &  9.77   & \\ 
                   &                                                            & Limitation of Statistical Weights  & 1.40(80)$\times$10$^{21}$  &  5.12   & \\ 
                   &                                                            & Normalized Residuals  & 1.60(77)$\times$10$^{21}$  &  3.84   & \\   
                   &                                                            & Rajeval Technique  & 1.40(80)$\times$10$^{21}$  & 0.45    &   \\ 
                   &                                                            & Expected Value  & 1.40(80)$\times$10$^{21}$  &     & \\ 
                   &                                                            & Bootstrap  & 1.40(67)$\times$10$^{21}$  &  5.12   & \\ 
                   &                                                            & Mandel-Paule  & 1.40(100)$\times$10$^{21}$  &  5.12    &\\                            
\hline
$^{150}$Nd & 2$\beta^-$(0$^{+}_{1}$$\rightarrow$0$^{+}_{1}$) & Unweighted Average  & 1.160(370)$\times$10$^{19}$   &     & 1.160(370)$\times$10$^{19}$ \\      
                   &                                                            & Weighted Average  & 8.40(120)$\times$10$^{18}$  & 6.25   &  \\ 
                   &                                                            & Limitation of Statistical Weights  & 8.20(130)$\times$10$^{18}$  & 5.86   &  \\ 
                   &                                                            & Normalized Residuals  & 8.70(130)$\times$10$^{18}$  & 4.68   &  \\   
                   &                                                            & Rajeval Technique  & 9.00(120)$\times$10$^{18}$  &  0.99   &  \\ 
                   &                                                            & Expected Value  & 8.80(290)$\times$10$^{18}$  &    &  \\ 
                   &                                                            & Bootstrap  & 1.100(480)$\times$10$^{19}$  &  19.37   & \\ 
                   &                                                            & Mandel-Paule  & 9.80(450)$\times$10$^{18}$  &  10.30   & \\    
\hline
$^{150}$Nd & 2$\beta^-$(0$^{+}_{1}$$\rightarrow$0$^{+}_{2}$) & Unweighted Average  & 1.12(17)$\times$10$^{20}$   &     & \\      
                   &                                                            & Weighted Average  & 1.17$^{+16}_{-15}$$\times$10$^{20}$  &  0.23   &  1.17$^{+26}_{-21}$$\times$10$^{20}$  \\ 
                   &                                                            & Limitation of Statistical Weights  & 1.17$^{+16}_{-15}$$\times$10$^{20}$  &  0.23   & \\ 
                   &                                                            & Normalized Residuals  & 1.17$^{+16}_{-15}$$\times$10$^{20}$  &  0.23   & \\   
                   &                                                            & Rajeval Technique  & 1.17$^{+16}_{-15}$$\times$10$^{20}$  & 0.23  &\\ 
                   &                                                            & Expected Value  & 1.12$^{+20}_{-13}$$\times$10$^{20}$  &     & \\ 
                   &                                                            & Bootstrap  & 1.21(20)$\times$10$^{20}$  &  0.26  &  \\ 
                   &                                                            & Mandel-Paule  & 1.12$^{+16}_{-15}$$\times$10$^{20}$  &  0.27    &\\    
\hline
$^{238}$U & 2$\beta^-$(0$^{+}_{1}$$\rightarrow$0$^{+}_{1}$) & Single Measurement  & 2.0(6)$\times10^{21}$  &    & 2.0(6)$\times10^{21}$ \\                                        
\hline      
\hline        
\label{Recom}
\end{longtable}
\end{landscape}

\subsection{Summary of Evaluated Half-Lives}

The collection of recently evaluated half-lives is shown in Table~\ref{Recent}. The current values can be compared  with ENSDF values~\cite{ensdf,Dim20,NSDD}, works of  Barabash (ITEP 2020)~\cite{Bar20,ASBar10}, and 2013 NNDC evaluation~\cite{Pri14}.  The present work capitalizes on the 2013 NNDC evaluation and supersedes it. Several other independent data evaluations are discussed below.

The ENSDF library contains complete nuclear data sets for all known nuclei (vertical evaluations), and it is the standard nuclear structure and decay library worldwide. The periodicity of ENSDF mass chain re-evaluations is 10-15 years and the evaluation updates for $^{130}$Te, and $^{238}$U double-beta half-lives are in the pipeline. The assessments of different ENSDF nuclei are performed by independent data scientists. This ensures the overall library consistency and protects the library against evaluation bias issues~\cite{Pri21}. At the same time, slightly different approaches and completion dates introduce additional uncertainties that may affect nuclear systematics. 

The nuclear systematics and periodicity arguments provide a preference for a comprehensive evaluation of nuclear properties across the nuclear chart (horizontal evaluations) in several cases. Such evaluations have been performed at NNDC and ITEP.

The ITEP works by Barabash satisfy many horizontal evaluation criteria while several data sets are ignored and the findings were somewhat misinterpreted. In multiple cases, Barabash analyzes the data and draws conclusions on precise double-beta decay T$_{1/2}$ values and nuclear matrix elements, however, the experimental findings speak otherwise. Many present measurements and experimental techniques are imprecise because experiments are not complete,  published results are often in disagreement and do not converge towards final values~\cite{Pri15},  and experimental two-neutrino mode half-lives change over time.  As a result, multiple evaluation procedures produce slightly different results, and data evaluators are forced to make difficult choices and introduce larger uncertainties. 

On the contrary, the ITEP uncertainties~\cite{Bar20} are relatively low. A simple analysis of the ITEP evaluation Table 1 data suggests that the author accepted uncertainties of the Particle Data Group (PDG)~\cite{PDG} averaging code. The recommended ITEP half-life uncertainties are slightly and substantially lower than experimental for $^{48}$Ca, $^{76}$Ge,  $^{116}$Cd and  $^{100}$Mo to the first excited $0^{+}$, $^{128}$Te decays,  respectively. It is a well-known issue in the nuclear data field that many computer codes produce unrealistic uncertainties, and the PDG code produces good results for statistical uncertainties. However, systematic uncertainties are dominant in many cases. The current situation was perfectly summarized by Donald Rumsfeld at the U.S. Defense Department meeting in February of 2002: "There are known knowns; there are things we know we know. We also know there are known unknowns; that is to say, we know there are some things we do not know. But there are also unknown unknown - the ones we don't know we don't know."  It is very difficult to evaluate the unrecognized sources
of uncertainty (USU), and the International Atomic Energy Agency (IAEA) coordinated research project~\cite{Cap20} provided multiple recommendations for evaluators. In light of this disclosure, data evaluators cannot accept smaller uncertainties than the lowest in the dataset that is being averaged. 

In conclusion, at the present state of development of nuclear science, we cannot deduce precise double-beta decay T$_{1/2}$ values. We can only produce realistic (most probable) values of two-neutrino mode half-lives for a particular date (literature time cut). These values have to be treated as evolving with time because of the constant maturation of experimental methods and techniques.

\begin{landscape}
\begin{longtable}[c]{l c| c| c| c |c} 
\caption{Recent recommended values of $2 \beta$(2$\nu$)-decay half-lives. Ground-to-ground state transitions are assumed unless daughter nucleus $\gamma$-rays were measured.} \\
\hline\hline
   \textbf{Nuclide}  &   \textbf{Decay} &   \textbf{Present Work} &  \textbf{ENSDF} &   \textbf{ITEP 2020 ~\cite{Bar20}} &  \textbf{NNDC 2013~\cite{Pri14}} \\
    \textbf{}  &   \textbf{} &   \textbf{T$_{1/2}$ (years)} &  \textbf{T$_{1/2}$ (years)} &   \textbf{T$_{1/2}$ (years)} &  \textbf{T$_{1/2}$  (years)} \\
 \hline 
\endfirsthead
\hline
   \textbf{Nuclide}  &   \textbf{Decay} &   \textbf{Present Work} &  \textbf{ENSDF} &   \textbf{ITEP 2020~\cite{Bar20}} &  \textbf{NNDC 2013~cire{Pri14}} \\
      \textbf{}  &   \textbf{} &   \textbf{(T$_{1/2}$  years)} &  \textbf{(T$_{1/2}$  years)} &   \textbf{(T$_{1/2}$  years)} &  \textbf{T$_{1/2}$  (years)} \\
 \hline 
\endhead
\hline
\multicolumn{6}{r}{\textit{Continued on next page ...}} \\
\endfoot
\endlastfoot
$^{48}$Ca & 2$\beta^-$(0$^{+}_{1}$$\rightarrow$0$^{+}_{1}$) & 5.960$^{+1389}_{-1082}$$\times$10$^{19}$  & 2.9$^{+42}_{-11}$$\times$10$^{19}$~\cite{2022CH04}  &  5.3$^{+12}_{-8}$$\times$10$^{19}$  & 4.39(58)$\times$10$^{19}$ \\      
\hline  
$^{76}$Ge & 2$\beta^-$(0$^{+}_{1}$$\rightarrow$0$^{+}_{1}$) &  1.745(89)$\times$10$^{21}$   &  1.926(94)$\times$10$^{21}$~\cite{2024SI02}  & 1.88(8)$\times$10$^{21}$ &  1.43(53)$\times$10$^{21}$ 	\\       
\hline
$^{82}$Se & 2$\beta^-$(0$^{+}_{1}$$\rightarrow$0$^{+}_{1}$) & 9.16(45)$\times$10$^{19}$ & 9.6(10)$\times$10$^{19}$~\cite{2019TU06}    & 8.7$^{+2}_{-1}$$\times$10$^{19}$  & 9.19(76)$\times$10$^{19}$  \\      
\hline  
$^{78}$Kr &  2$K$(0$^{+}_{1}$$\rightarrow$0$^{+}_{1}$) & 1.90$^{+133}_{-76}\times10^{22}$  &  $\geq$1.5$\times$10$^{21}$~\cite{2009FA09}  & 1.9$^{+13}_{-8}$$\times$10$^{22}$  &  \\      
\hline  
$^{96}$Zr & 2$\beta^-$(0$^{+}_{1}$$\rightarrow$0$^{+}_{1}$) & 2.335(210)$\times$10$^{19}$ & 2.0(4)$\times$10$^{19}$~\cite{2008AB19}  &  2.3(2)$\times$10$^{19}$  & 2.16(26)$\times$10$^{19}$ \\      
\hline  
$^{100}$Mo & 2$\beta^-$(0$^{+}_{1}$$\rightarrow$0$^{+}_{1}$) & 7.08(14)$\times$10$^{18}$  & 7.01$^{+21}_{-17}$$\times$10$^{18}$~\cite{2021SI08}   &  7.06$^{+15}_{-13}$$\times$10$^{18}$   &  6.98(44)$\times$10$^{18}$ \\      
\hline  
$^{100}$Mo & 2$\beta^-$(0$^{+}_{1}$$\rightarrow$0$^{+}_{2}$) & 6.92(85)$\times$10$^{20}$  & 6.9(9)$\times$10$^{20}$~\cite{2021SI08}  &  6.7$^{+5}_{-4}$$\times$10$^{20}$  & 5.70(136)$\times$10$^{20}$ 	 \\      
\hline  
$^{116}$Cd & 2$\beta^-$(0$^{+}_{1}$$\rightarrow$0$^{+}_{1}$) & 2.738(120)$\times$10$^{19}$  &  3.3(4)$\times$10$^{19}$~\cite{2010BL02}  &   2.69(9)$\times$10$^{19}$ & 2.89(25)$\times$10$^{19}$ \\      
\hline   
$^{128}$Te & 2$\beta^-$(0$^{+}_{1}$$\rightarrow$0$^{+}_{1}$) &  2.167(200)$\times$10$^{24}$   & 7.7(4)$\times$10$^{24}$~\cite{2015EL04}   &  2.25(9)$\times$10$^{24}$     & 3.49(199)$\times$10$^{24}$ 	\\                                                                                                                  
\hline
$^{130}$Te & 2$\beta^-$(0$^{+}_{1}$$\rightarrow$0$^{+}_{1}$) & 8.69(18)$\times$10$^{20}$  &  $>$7.9$\times$10$^{20}$~\cite{2001SI26} &  7.91(21)$\times$10$^{20}$   & 7.14(104)$\times$10$^{20}$ \\       
\hline  
$^{124}$Xe &  2$\epsilon$(0$^{+}_{1}$$\rightarrow$0$^{+}_{1}$) & 1.10(22)$\times10^{22}$ &  $\geq$1.6$\times$10$^{14}$~\cite{2008KA21}  &  1.8(5)$\times$10$^{22}$   & \\                     
\hline
$^{136}$Xe & 2$\beta^-$(0$^{+}_{1}$$\rightarrow$0$^{+}_{1}$) &  2.240(61)$\times$10$^{21}$ &  2.165(61)$\times$10$^{21}$~\cite{2018MC07}  & 2.18(5)$\times$10$^{21}$  & 2.34(13)$\times$10$^{21}$ \\      
\hline
$^{130}$Ba &  2$\epsilon$(0$^{+}_{1}$$\rightarrow$0$^{+}_{1}$) & 1.40(80)$\times$10$^{21}$ &    & 2.2(5)$\times$10$^{21}$ & 1.40(80)$\times$10$^{21}$ 	\\                             
\hline
$^{150}$Nd & 2$\beta^-$(0$^{+}_{1}$$\rightarrow$0$^{+}_{1}$) & 1.160(370)$\times$10$^{19}$  &  0.91(7)$\times$10$^{19}$~\cite{2013BA31}  &  9.34(65)$\times$10$^{18}$ & 8.37(45)$\times$10$^{18}$ \\         
\hline
$^{150}$Nd & 2$\beta^-$(0$^{+}_{1}$$\rightarrow$0$^{+}_{2}$) & 1.17$^{+26}_{-21}$$\times$10$^{20}$  &  1.33$\times$10$^{20}$~\cite{2013BA31}   &   1.2$^{+3}_{-2}$$\times$10$^{20}$  & 1.33(40)$\times$10$^{20}$ \\       
\hline
$^{238}$U & 2$\beta^-$(0$^{+}_{1}$$\rightarrow$0$^{+}_{1}$) & 2.0(6)$\times10^{21}$ &   &  2.00(60)$\times$10$^{21}$ & 2.00(60)$\times$10$^{21}$ 	\\                                        
\hline      
\hline        
\label{Recent}
\end{longtable}
\end{landscape}

\section{Effective Nuclear Matrix Elements}

Double-beta decay nuclear matrix elements are not directly observed quantities, they can be deduced from experimental or evaluated half-lives. 
 T$_{1/2}^{2\nu}$ values are often described as a product of phase space factors and nuclear matrix elements~\cite{19Sto}
\begin{equation}
\label{myeq.Half1}  
(T_{1/2}^{2\nu} (0^{+} \rightarrow 0^{+}))^{-1} = G^{2 \nu} (E,Z) \times g_{A}^{4} \times |m_{e} c^{2} M^{2 \nu}|^{2}  = G^{2 \nu} (E,Z) \times |M^{2 \nu}_{eff}|^{2}, 
\end{equation}
where the function  $G^{2 \nu} (E,Z)$ results from lepton phase space integration and contains all relevant constants, g$_{A}$ is the axial vector
coupling constant  and 
$|M^{2 \nu}|$  and $|M^{2 \nu}_{eff}|^{2}$ are nuclear matrix and effective nuclear matrix elements, respectively. The calculations of nuclear matrix elements are not trivial and results may vary, while phase space factors are better understood.  Using the present work evaluated half-lives and theoretical phase space factors one can deduce reliable values of effective nuclear matrix elements.

\subsection{Compilation of Phase Space Factors}
\begin{figure}[ht]
 \centering
\includegraphics[width=0.8\textwidth]{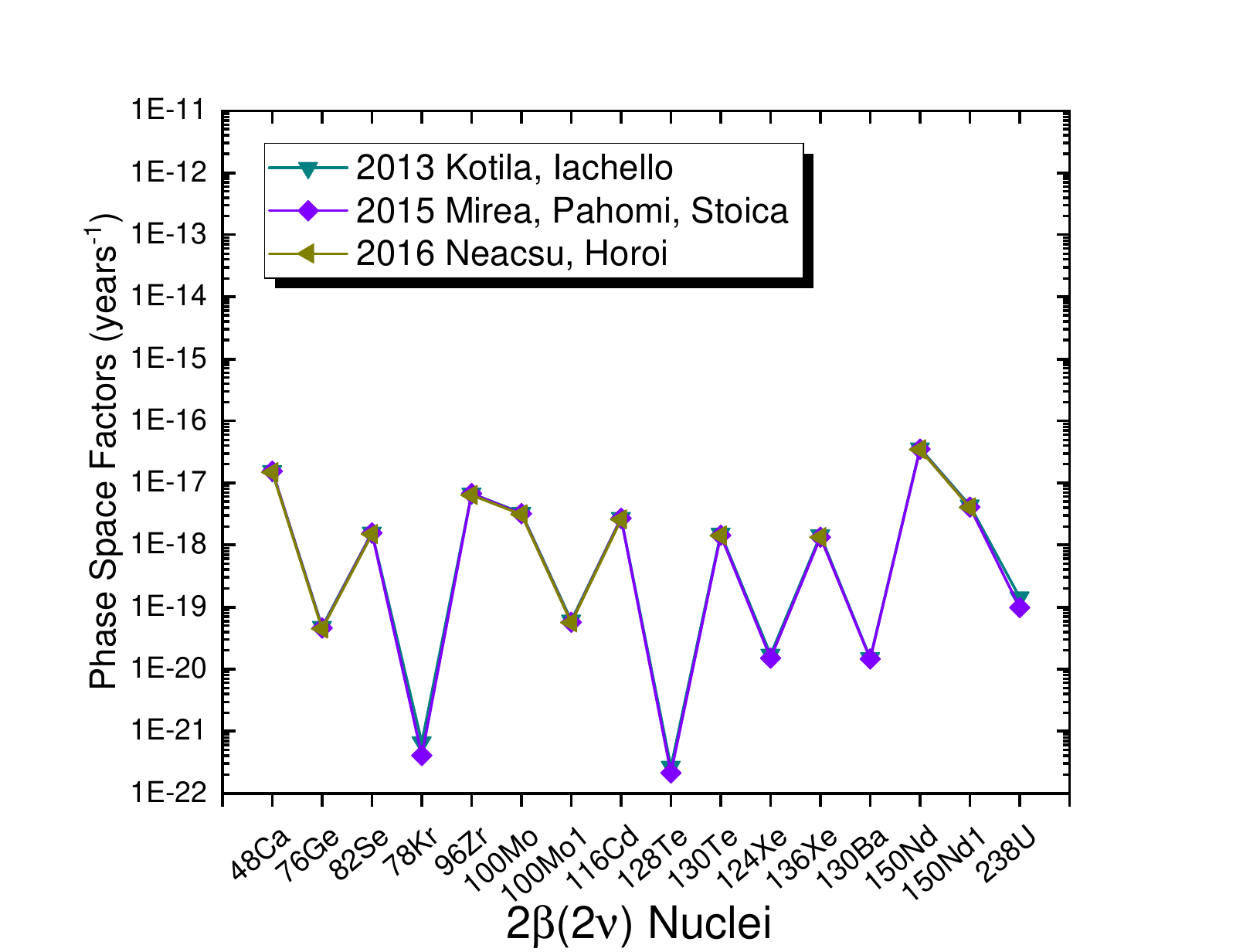}
\includegraphics[width=0.8\textwidth]{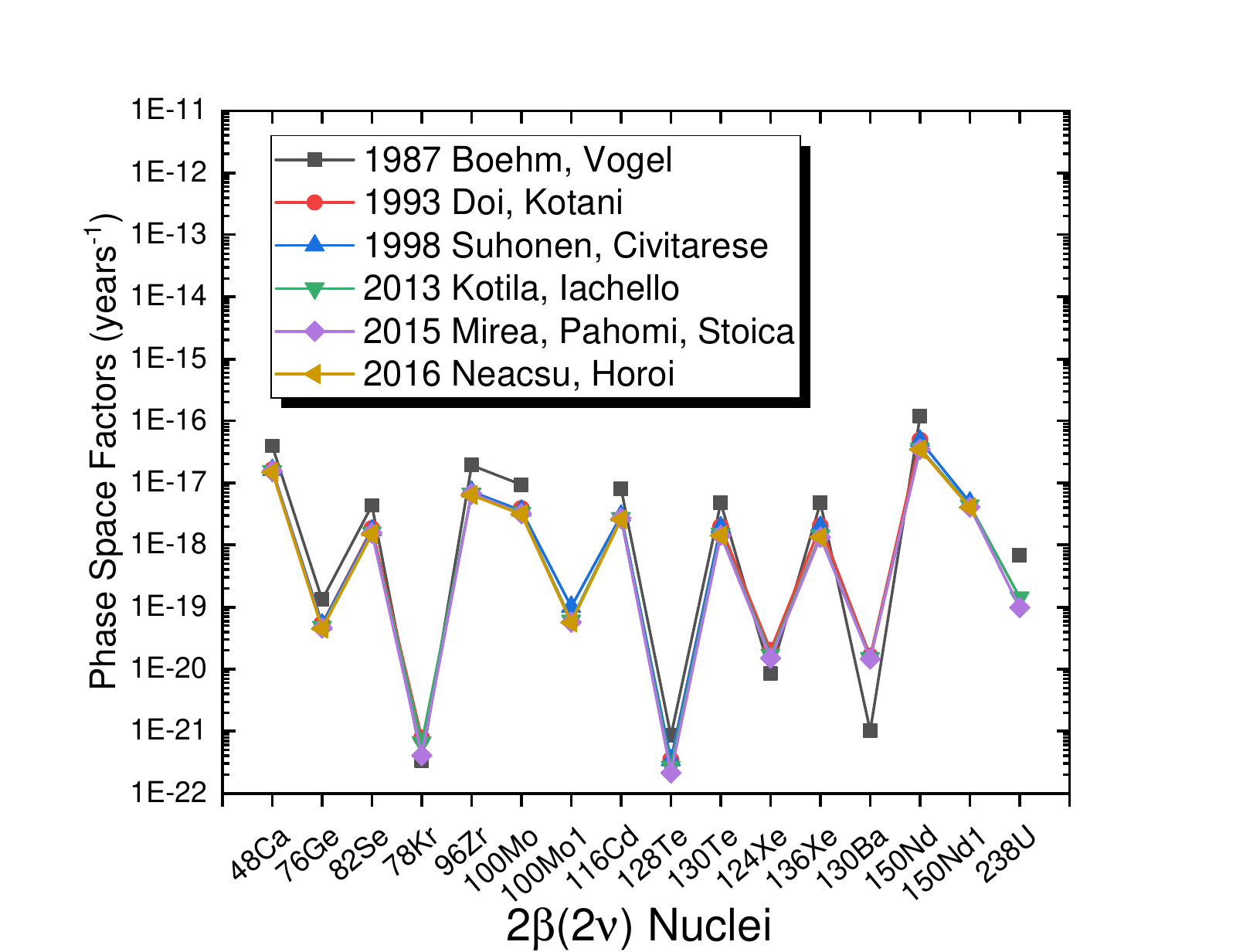}
\caption{Phase Space Factors: Six historical~\cite{87Bo,85Doi,92Doi,98Suh,12Kot,24Kot,13Kot,15Mir,16Nea} and three recent~\cite{12Kot,24Kot,13Kot,15Mir,16Nea} calculations are shown 
 in the bottom and top panels, respectively.}
\label{fig:PhaseFactors}
\end{figure}

Phase space factors for nuclei of interest are compiled in Table~\ref{RecentPF} and Fig.~\ref{fig:PhaseFactors}.  Further details could be found in a recent review~\cite{19Sto}. 

{\small
\begin{landscape}
\begin{longtable}[c]{l c| c| c| c |c |c |c} 
\caption{Theoretical phase space factors of $2 \beta$(2$\nu$)-decay half-lives~\cite{87Bo,85Doi,92Doi,98Suh,12Kot,24Kot,13Kot,15Mir,16Nea}. In Ref.~\cite{87Bo}, values of $G^{2\nu} g_A^4$ were given, and $g_A=1.25$. 
In work~\cite{85Doi}, $G^{2\nu}$ included $g_A=1.25$, in  publication~\cite{92Doi}, $g_A=1.261$, and in paper~\cite{98Suh}, $g_A=1.254$. A comma character was used to separate groups of thousands, and a period character indicates the decimal place.} \\
\hline\hline
   \textbf{Nuclide}  &   \textbf{Decay} &   \textbf{$(G^{2\nu})$~\cite{87Bo}} &  \textbf{$(G^{2\nu})$~\cite{85Doi,92Doi}} & \textbf{$(G^{2\nu})$~\cite{98Suh} } &   \textbf{$(G^{2\nu})$~\cite{12Kot,24Kot,13Kot}} & \textbf{$(G^{2\nu})$~\cite{15Mir}}  & \textbf{$(G^{2\nu})$~\cite{16Nea}} \\
    \textbf{}  &   \textbf{} &   \textbf{ (years)$^{-1}$} &  \textbf{(years)$^{-1}$} &  \textbf{(years)$^{-1}$} &   \textbf{(years)$^{-1}$} &  \textbf{(years)$^{-1}$} &  \textbf{(years)$^{-1}$} \\
 \hline 
\endfirsthead
\hline
   \textbf{Nuclide}  &   \textbf{Decay} &   \textbf{$(G^{2\nu})$~\cite{87Bo}} &  \textbf{$(G^{2\nu})$~\cite{85Doi,92Doi,93Doi}}  &  \textbf{$(G^{2\nu})$~\cite{98Suh} } &   \textbf{$(G^{2\nu})$~\cite{12Kot,24Kot,13Kot}} & \textbf{$(G^{2\nu})$~\cite{15Mir}}  & \textbf{$(G^{2\nu})$~\cite{16Nea}} \\
      \textbf{}  &   \textbf{} &   \textbf{(years)} &  \textbf{(years)} &  \textbf{((years)$^{-1}$)} &   \textbf{(years)$^{-1}$} &  \textbf{(years)$^{-1}$} &  \textbf{(years)$^{-1}$} \\
 \hline 
\endhead
\hline
\multicolumn{6}{r}{\textit{Continued on next page ...}} \\
\endfoot
\endlastfoot
$^{48}$Ca  & 2$\beta^-$(0$^{+}_{1}$$\rightarrow$0$^{+}_{1}$) &   16,254E-21 & 16,249E-21 &16,176E-21  & 15,550E-21 & 15,536E-21 & 14,804.6E-21	\\       
\hline
$^{76}$Ge & 2$\beta^-$(0$^{+}_{1}$$\rightarrow$0$^{+}_{1}$) &  53.47E-21 & 53.78E-21 & 52.57E-21  & 48.1E-21 & 46.47E-21 & 45.1E-21	\\       
\hline
$^{82}$Se & 2$\beta^-$(0$^{+}_{1}$$\rightarrow$0$^{+}_{1}$) & 1,781E-21 & 1,782E-21 & 1,739E-21  & 1,590E-21 & 1,573E-21 &1,503.1E-21 \\      
\hline  
$^{78}$Kr &  2$\epsilon$(0$^{+}_{1}$$\rightarrow$0$^{+}_{1}$) &  0.1365E-21  &  0.774E-21  & 0.8088E-21  & 0.660E-21 & 410E-24 & \\      
\hline  
$^{96}$Zr & 2$\beta^-$(0$^{+}_{1}$$\rightarrow$0$^{+}_{1}$) & 7,892E-21  &  & 7,279E-21  & 6,810E-21 & 6,744E-21 & 6,420E-21 \\      
\hline  
$^{100}$Mo & 2$\beta^-$(0$^{+}_{1}$$\rightarrow$0$^{+}_{1}$) & 3,864E-21  & 3,864E-21 & 3,599E-21   &   3,310E-21 & 3,231E-21 & 3,106E-21 \\      
\hline  
$^{100}$Mo & 2$\beta^-$(0$^{+}_{1}$$\rightarrow$0$^{+}_{2}$) &   &  & 101E-21    & 60.55E-21  & 57.08E-21 &  56.208E-21	 \\      
\hline  
$^{116}$Cd & 2$\beta^-$(0$^{+}_{1}$$\rightarrow$0$^{+}_{1}$) & 3,277E-21 &  & 2,993E-21  & 2,770E-21  & 2,688E-21 & 2,5878E-21 \\      
\hline   
$^{128}$Te & 2$\beta^-$(0$^{+}_{1}$$\rightarrow$0$^{+}_{1}$) & 0.3471E-21  & 0.3498E-21 & 0.3437E-21 &  0.269E-21 & 0.2149E-21 & 	\\                                                                                                                  
\hline
$^{130}$Te & 2$\beta^-$(0$^{+}_{1}$$\rightarrow$0$^{+}_{1}$) & 1,969E-21  & 1,967E-21 & 1,941E-21 & 1,530E-21 & 1,442E-21  & 1,427.3E-21 \\       
\hline  
$^{124}$Xe &  2$\epsilon$(0$^{+}_{1}$$\rightarrow$0$^{+}_{1}$) &  3.501E-21 & 20.17E-21 & 20.62E-21  & 17.200E-21  & 15.096E-21 &  \\                     
\hline
$^{136}$Xe & 2$\beta^-$(0$^{+}_{1}$$\rightarrow$0$^{+}_{1}$) &  1,979E-21 & 1,990E-21 & 1,982E-21  & 1,430E-21 & 1,332E-21 & 1,337.3E-21 \\      
\hline
$^{130}$Ba &  2$\epsilon$(0$^{+}_{1}$$\rightarrow$0$^{+}_{1}$) & 2.592E-21  &  16.35E-21 & 16.58E-21 & 15.000E-21 &	14.773E-21 & \\                             
\hline
$^{150}$Nd & 2$\beta^-$(0$^{+}_{1}$$\rightarrow$0$^{+}_{1}$) & 48,704E-21 & 48,701E-21 & 48,528E-21  & 36,400E-21  & 35,397E-21  & 34,675.3E-21 \\         
\hline
$^{150}$Nd & 2$\beta^-$(0$^{+}_{1}$$\rightarrow$0$^{+}_{2}$) &   &  & 4,852E-21 &  4,329E-21   & 4,116E-21 & 4,063.75E-21 \\       
\hline
$^{238}$U & 2$\beta^-$(0$^{+}_{1}$$\rightarrow$0$^{+}_{1}$) & 278.6E-21 &  & & 144E-21  & 98.51E-21 & 	\\                                        
\hline      
\hline        
\label{RecentPF}
\end{longtable}
\end{landscape}

} 

\subsection{Effective Nuclear Matrix Element Values and Ranges}

Two-neutrino mode half-lives are products of phase space factors and effective nuclear matrix elements as shown in Eq.~\ref{myeq.Half1}. Using Table~\ref{Recent} recommended half-lives and Table~\ref{RecentPF} phase space factors we deduce effective nuclear matrix element values and ranges. The matrix elements uncertainties are calculated as derivatives of Eq.~\ref{myeq.Half1}
\begin{equation}
\label{myeq.Unc}  
d|M^{2 \nu}_{eff}| = -G^{2 \nu} (E,Z) \times |M^{2 \nu}_{eff}|^{3} \times dT_{1/2}^{2\nu} (0^{+} \rightarrow 0^{+})/2,
\end{equation}
or
\begin{equation}
\label{myeq.Unc1}  
d|M^{2 \nu}_{eff}| = -|M^{2 \nu}_{eff}| \times dT_{1/2}^{2\nu} (0^{+} \rightarrow 0^{+})/(2 \times T_{1/2}^{2\nu} (0^{+} \rightarrow 0^{+})),
\end{equation}
where $G^{2 \nu} (E,Z)$, $|M^{2 \nu}_{eff}|$ and $dT_{1/2}^{2\nu} (0^{+} \rightarrow 0^{+})$ are theoretical phase space factors, deduced effective nuclear matrix elements and evaluated half-life uncertainties, respectively.

The shown in Table~\ref{RecentPF} phase space factors agree within 10-20$\%$, and their numerical values are decreasing in each subsequent calculation~\cite{19Sto}. This implies a lack of convergence or ranges for effective nuclear matrix elements. To deduce nuclear matrix element ranges we will omit the three older phase space factor calculations~\cite{87Bo,85Doi,92Doi,98Suh}, and select values computed in the last ten years~\cite{12Kot,13Kot,15Mir,16Nea}. The resulting matrix elements could be averaged using the present work procedures. 

The present work's effective nuclear matrix elements and averages are shown in Table~\ref{RecentNME}. These results can be compared with the NNDC~\cite{Pri14} and ITEP~\cite{Bar20} values. The NNDC matrix elements were based on the best phase space factor calculation available at that time~\cite{12Kot,13Kot} while the 2020 ITEP values have incorporated another calculation~\cite{12Kot,13Kot,15Mir} to produce recommended values. The current work unweighted averages of three-phase space factor calculations are consistent with the ITEP recommendations and explain the disagreement of $^{238}$U matrix elements between NNDC and ITEP. Simultaneously the current work findings highlight the urgent need for convergence of phase space factor calculations and improved precision.

{\small
\begin{landscape}
\begin{longtable}[c]{l c| c| c| c |c |c |c} 
\caption{Effective Nuclear Matrix Elements for $2 \beta$(2$\nu$)-decay. Third, fourth, and fifth column values are based on the present work half-lives and theoretical phase space factors~\cite{12Kot,24Kot,13Kot,15Mir,16Nea}. The sixth column shows the unweighted average for the previous three columns. These data are compared with the previous results~\cite{Pri14,Bar20}.} \\
\hline\hline
   \textbf{Nuclide}  &   \textbf{Decay} &   \textbf{$|M^{2 \nu}_{eff}|$~\cite{12Kot,24Kot,13Kot}} &  \textbf{$|M^{2 \nu}_{eff}|$~\cite{15Mir}} & \textbf{$|M^{2 \nu}_{eff}|$~\cite{16Nea}}  &   \textbf{$|M^{2 \nu}_{eff}|$ Average} & \textbf{$|M^{2 \nu}_{eff}|$~\cite{Pri14}}  & \textbf{$|M^{2 \nu}_{eff}|$~\cite{Bar20}} \\
    \textbf{}  &   \textbf{} &   \textbf{Present Work} &  \textbf{Present Work} &  \textbf{Present Work} &   \textbf{Present Work} &  \textbf{} &  \textbf{} \\
 \hline 
\endfirsthead
\hline
   \textbf{Nuclide}  &   \textbf{Decay} &   \textbf{$|M^{2 \nu}_{eff}|$~\cite{12Kot,13Kot}} &  \textbf{$|M^{2 \nu}_{eff}|$~\cite{15Miri}}  &  \textbf{$|M^{2 \nu}_{eff}|$~\cite{16Nea}} &   \textbf{$|M^{2 \nu}_{eff}|$ Average} & \textbf{$|M^{2 \nu}_{eff}|$~\cite{Pri14}}  & \textbf{$|M^{2 \nu}_{eff}|$~\cite{Bar20}} \\
      \textbf{}  &   \textbf{} &   \textbf{Present Work} &  \textbf{Present Work} &  \textbf{Present Work} &   \textbf{Present Work} &  \textbf{} &  \textbf{} \\
 \hline 
\endhead
\hline
\multicolumn{6}{r}{\textit{Continued on next page ...}} \\
\endfoot
\endlastfoot
$^{48}$Ca  & 2$\beta^-$(0$^{+}_{1}$$\rightarrow$0$^{+}_{1}$) &  0.03285$_{-383}^{+298}$  &  0.03286$_{-383}^{+298}$  &  0.03366$_{-392}^{+306}$  &  0.03312$^{+300}_{-380}$  & 0.0383(25) &  0.035(3)	\\       
\hline
$^{76}$Ge & 2$\beta^-$(0$^{+}_{1}$$\rightarrow$0$^{+}_{1}$) &   0.10915(278) &  0.11105(283)  &  0.11272(287)   &  0.11090(278)  & 0.120(21) &  0.106(4)	\\       
\hline
$^{82}$Se & 2$\beta^-$(0$^{+}_{1}$$\rightarrow$0$^{+}_{1}$) &  0.08286(204)  &  0.08331(205) &  0.08522(209)  &  0.08370(204)  & 0.0826(34) & 0.085(1) \\      
\hline  
$^{78}$Kr &  2$\epsilon$(0$^{+}_{1}$$\rightarrow$0$^{+}_{1}$) &   0.28239$_{-9884}^{+5648}$   &  0.35829$_{-12540}^{+7166}$  &  &  0.32000$^{+5600}_{-9900}$ &    & 0.32$^{+15}_{-11}$ \\      
\hline  
$^{96}$Zr & 2$\beta^-$(0$^{+}_{1}$$\rightarrow$0$^{+}_{1}$) &  0.07930(357)  &  0.07969(358)  &  0.08167(367)   &  0.08021(360)  & 0.0824(50) &  0.080(4) \\      
\hline  
$^{100}$Mo & 2$\beta^-$(0$^{+}_{1}$$\rightarrow$0$^{+}_{1}$) &  0.20657(204)  &  0.20908(207)  &    0.21325(211)  &  0.20970(204)   & 0.208(7) & 0.185(2) \\      
\hline  
$^{100}$Mo & 2$\beta^-$(0$^{+}_{1}$$\rightarrow$0$^{+}_{2}$) &  0.15449(949)   &  0.15911(977)  &    0.16034(985)  &  0.15800(950)  & 0.170(20) & 0.151(5)  \\      
\hline  
$^{116}$Cd & 2$\beta^-$(0$^{+}_{1}$$\rightarrow$0$^{+}_{1}$) &  0.11483(252)  &  0.11657(255)  &   0.11880(260)  &   0.11680(252)  & 0.112(5) & 0.108(3) \\      
\hline   
$^{128}$Te & 2$\beta^-$(0$^{+}_{1}$$\rightarrow$0$^{+}_{1}$) &  0.04143(191)   &  0.04634(214)  &   &  0.04390(191)  & 0.0326(93)  & 0.043(3)	\\                                                                                                                  
\hline
$^{130}$Te & 2$\beta^-$(0$^{+}_{1}$$\rightarrow$0$^{+}_{1}$) &  0.02742(28)  &  0.02825(29)  &  0.02839(29) &  0.02802(28)  & 0.0303(22)  & 0.0293(9) \\       
\hline  
$^{124}$Xe &  2$\epsilon$(0$^{+}_{1}$$\rightarrow$0$^{+}_{1}$) &   0.07270(727)  &  0.07760(776)  &   &  0.07520(727)   &  & 0.059$^{+13}_{-9}$ \\                     
\hline
$^{136}$Xe & 2$\beta^-$(0$^{+}_{1}$$\rightarrow$0$^{+}_{1}$) &  0.01767(24)  &  0.01831(25)  &  0.01827(25)  &  0.01808(24)  & 0.0173(5) & 0.0181(6)  \\      
\hline
$^{130}$Ba &  2$\epsilon$(0$^{+}_{1}$$\rightarrow$0$^{+}_{1}$) &  0.21822(6235)  &   0.21989(6283)  &  &  0.21906(6235)  &	0.218(62) & 0.175$^{+24}_{-17}$ \\                             
\hline
$^{150}$Nd & 2$\beta^-$(0$^{+}_{1}$$\rightarrow$0$^{+}_{1}$) &   0.04867(776)  &  0.04940(790)  &  0.04990(800)   &   0.04930(776) & 0.0572(15)  & 0.055(3) \\         
\hline
$^{150}$Nd & 2$\beta^-$(0$^{+}_{1}$$\rightarrow$0$^{+}_{2}$) &  0.04443$^{+494}_{-399}$   &  0.04557$^{+506}_{-409}$  &  0.04586$^{+510}_{-412}$   &    0.04530$^{+494}_{-399}$   & 0.0417(63)  & 0.044(5) \\       
\hline
$^{238}$U & 2$\beta^-$(0$^{+}_{1}$$\rightarrow$0$^{+}_{1}$) &  0.05893(884)  &  0.07124(1069)  &    &  0.06510(884)   & 0.185(28) & 0.13$^{+9}_{-7}$	\\                                        
\hline      
\hline        
\label{RecentNME}
\end{longtable}
\end{landscape}

} 

\section{Conclusions}

The double-beta decay observations were compiled and analyzed thoroughly. The recommended half-lives were deduced using the standard USNDP procedures and techniques. The final results are the best available at the present day. They reflect the rare nature of this process, small statistics, issues with current experimental techniques enriched isotope production, and detector technologies. 

In evaluation, we concentrated on half-lives because they are measured quantities or observables in double-beta decay. The recommended half-lives reflect the nuclear structure and decay properties of parent, intermediate, and daughter nuclei. These properties are encapsulated in nuclear matrix elements and phase space factors. Using the presently available calculations of phase space factors,  the effective nuclear matrix elements were deduced and compared with other results. The analysis is impacted by the potential issues with phase space factor calculations that reduce the precision of extracted nuclear matrix elements.

The present work is a comprehensive evaluation of double-beta decay half-lives and nuclear matrix elements that is based on all available data and state of an art data analysis techniques. A large number of theoretical and experimental works have been dedicated to these topics in the last 85 years, and we contribute a set of independent benchmarks for further studies. 

\section{Acknowledgments}
\label{sec:Acknowledgements}
The authors are indebted to the late Dr. Balraj Singh (1941-2023)~\cite{24Dim} for his love of nuclear data evaluation and compilation work, positive influence, and sharing of best data evaluation practices that were acquired over the long and productive career, and Michael Birch (McMaster University) for developing nuclear data evaluation tools and procedures. 
We recognize Dr. J. Kotila (Yale University) and the referee for good comments and productive suggestions that helped to improve the manuscript. 
Work at Brookhaven was funded by the Office of Nuclear Physics, Office of Science of the U.S. Department
of Energy, under Contract No. DE-SC0012704 with Brookhaven Science Associates, LLC.  \\ \\




\clearpage

\end{document}